\documentclass[prb,twocolumn,aps,10pt]{revtex4-2}
\usepackage{graphicx,amssymb,amsfonts,amsmath}
\usepackage{color}
\usepackage{mathtools}
\usepackage{latexsym}
\usepackage{graphicx}
\usepackage{float}
\usepackage{dcolumn}
\usepackage{bm}
\usepackage{amssymb}
\usepackage{amsmath}
\usepackage{mathtools}
\usepackage[space]{grffile}
\usepackage{xcolor}
\usepackage{comment}
\usepackage{bbold}

\usepackage{hyperref}
\hypersetup{
    colorlinks=true,
    linkcolor=blue,
    filecolor=blue,      
    urlcolor=blue,
    citecolor=blue
    }
\DeclareUnicodeCharacter{2212}{\ensuremath{-}}
\renewcommand{\vec}[1]{{\boldsymbol #1}}

\begin{document}
\title{Stability of Majorana modes in Coulomb-disordered topological insulator nanowires}
\author{Leonard Kaufhold}
\author{Achim Rosch}
\affiliation{
Institute for Theoretical Physics, University of Cologne, 50937 Cologne, Germany
}

\begin{abstract}
We evaluate theoretically the possibility to realize Majorana zero modes in hybrid devices made from topological-insulator (TI) nanowires proximity-coupled to a superconductor. Such systems have been suggested as building blocks of future topological quantum computers, as they  have been predicted to realize Majorana zero modes protected by large gaps. A main obstacle is, however, the presence of a relatively large density of charged impurities, $n_\text{imp}\sim 10^{19}$\,cm$^{-3}$. 
Based on extensive numerical simulations, we show that the proximity to the superconductor leads to an efficient screening of the disorder potential. By  analyzing the Majorana splitting energy, the size of the Andreev gap and the localization of edge modes,  we show that robust Majorana modes can be realized for realistic levels of impurity concentrations and wire radii.
\end{abstract}
\maketitle

\section{Introduction}

Although the theoretical postulate of the Majorana fermion as a real-valued solution of the Dirac equation \cite{Majorana1937} is approaching its centennial, its experimental detection in particle physics remains elusive.
In the field of solid-state physics however, Majorana excitations naturally occur, e.g., as edge modes of topological superconductors. In the core of magnetic vortices and at the boundary of one-dimensional topological superconductors, Majorana states emerge at zero energy \cite{Kitaev2001,FuKane2008,Oreg2010,Lutchyn2010,Qi2011,Beenakker2013,Alicea2012}. 
 Spatially separated Majorana zero modes (MZM) can be used to encode qubits, and such systems exhibit a number of properties, which make them intriguing building blocks for scalable quantum computing \cite{Kitaev2003,Nayak2008,Plugge_2016,
Karzig2017,Manousakis2017,MicrosoftRoadmap2025}. First, MZM enjoy topological protection from a wide class of local perturbations. Second, certain classes of quantum gates can be implemented in a topological way, exploiting their
 non-Abelian exchange statistics \cite{Ivanov2001,Kitaev2006,Beenakker2013}.
 
 Three ingredients are needed to realize non-degenerate MZM in one-dimensional systems: superconductivity, sufficiently strong spin-orbit coupling, and time-reversal symmetry breaking by either an external magnetic field or by magnetism.
 As has been pointed out by Oreg {\it et al.} and Lutchyn {\it et al.} \cite{Oreg2010,Lutchyn2010} a promising platform for such systems are semiconductor nanowires with strong spin-orbit coupling, such as InAs and InSb based devices, in contact to an ordinary s-wave superconductor. By virtue of the proximity effect \cite{Volkov1995,Recher2001,FuKane2008,Chang2015}, the semiconducting nanowire becomes effectively superconducting and, through the application of a  magnetic field, can be brought into a topological superconducting phase. Under these conditions MZM can be realized in such devices.
\begin{figure}[t]\label{fig:IntroSchematic}
\centering
\includegraphics[width=6cm]{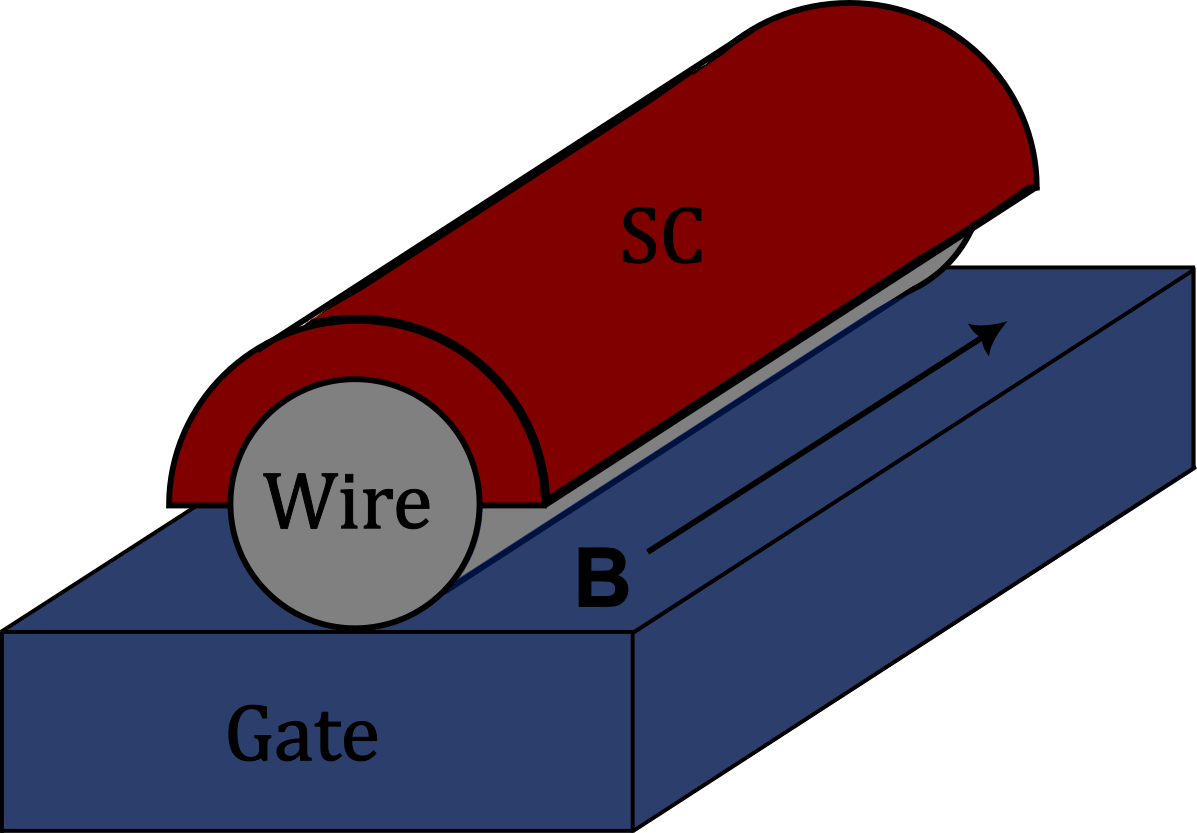}
\caption{Schematic model of the considered TI-SC hybrid nanowire. The system consists of a cylindrical wire of fixed radius 10-20nm and length 1$\mu$m close to a superconducting layer coating exactly half of the TI.}\label{fig:wire_schematic}
\end{figure}
This proposal attracted not only the interest of the scientific community \cite{Flensberg2010,Stoudenmire2011,Fidkowski2011,Mourik2012,Hamdou2013,Albrecht2016,Churchill2013,Deng2012}, but also commercial attention, most prominently by Microsoft
\cite{Suominen2017,Lutchyn2018,Microsoft2023,Microsoft2025}. 
The experimental observation of Majorana zero modes in semiconductor nanowires has made substantial progress showing promising results. An unambiguous detection of Majorana modes has, however, remained controversial over the years. 
Due to disorder effects  both the realization and detection of MZMs are confronted with major obstacles \cite{Bagrets2012,Liu2012,Sau2013,Kells2012,Neven2013,Prada2020,Hess2023,Pan2022,Yu2021,Legg2024,Legg2025b}.

An alternative platform to realize MZMs is based on the surface states of topological-insulator (TI) nanowires \cite{Cook2011}, which have seen significant progress in terms of material fabrication recently \cite{Arango2016,Breunig2022b,Muenning2021,Roessler2023}. Compared to the semiconductor nanowires, the TI nanowires have several major advantages \cite{Cook2011,Manousakis2017,Legg2021}: The surface states are strongly spin-orbit coupled and the energy-scales of spin orbit coupling are typically an order of magnitude larger than in semiconductor nanowires \cite{Cook2011,Ziegler2018,Prada2020,Legg2021}. Another advantage is the way in which time-reversal symmetry is broken. While semiconductor nanowires rely on the Zeemann effect (enhanced by orbital effects), TI nanowires use instead that a magnetic flux inside the wire affects
the surface states via the Aharonov-Bohm effect. It has been pointed out by Legg {\it et al.} \cite{Legg2021} that using gate voltages in this system allows to realize MZMs even with relatively small magnetic fluxes threading the nanowire. 
The level of disorder and especially the density of charged impurities may, however, be substantially larger in bulk-insulating topological insulators compared to semiconductors.

In topological insulators, the presence of a small density of charged impurities on the level of a few ppm cannot be avoided. Bulk-insulating samples are obtained by compensation doping to achieve the same density of positively and negatively charged defects.
In such compensated bulk-insulating topological insulators charged impurities create locally conducting regions, so-called electron-hole puddles \cite{Skinner2012,Huang2021}, which can, e.g., be measured using optical conductivity experiments \cite{Borgwardt2016}. Charge puddles can explain the giant magnetoresistance observed in Ref.~\cite{Breunig2017} that has been observed in STM experiments \cite{Knispel2017}, and limit the size of currents in quantum-anomalous Hall experiments \cite{Lippertz2022}. The magnitude of these effects are consistent with a density of charged impurities of the order of $10^{19}\,\text{cm}^{-3}$ in the investigated Bi-based TIs but the precise values will depend on details of the materials' fabrication.

The central question addressed in this paper is whether robust Majorana zero modes realistically exist in the presence of charged defects. We therefore develop a fully quantitative theory of MZMs in the presence of realistic densities of of charged impurities. Refs. \cite{burke2024,Heffels2022} have also investigated how disorder affects MZMs in hybrid TI nanowires, but the authors focus on a phenomenological model of Gaussian disorder. The problem was of charged impuirities was analyzed analytically by Huang and Shklovskii \cite{Huang2021}, which in their conclusions also point out the importance of screening by the superconductor.

In the following, we first discuss in Section \ref{secB} the numerical challenges and how we can derive a model able to quantitatively describe superconducting TI surfaces as well as the edge states in the presence of disorder by charged impurities. 
In Sec. \ref{secC}, we will explore screening effects by electron-electron interactions and by the superconductor. 
In Sec. \ref{secD} we  statistically analyze the impact of varying disorder strength on the stability of MZM.

\section{Model and simulation strategy}\label{secB}

Our goal is to analyze quantitatively the role of charged impurities on TI nanowires for experimentally realistic parameters.
The devices we have in mind are nanowires with a length of the order of $1\,\mu$m and with a radius of up to 20\,nm with a density of charged impurities in the range of $10^{19}$\,cm$^{-3}$. We therefore have to consider systems of $\mathcal{O}(10^9)$ atoms with $\mathcal{O}(10^4)$ charged impurities. A brute force  microscopic modeling and exact diagonalization of such systems is not possible, even when interactions are ignored. 
Furthermore, one cannot simply average diagrammatically over disorder configuration 
as in-gap states arise from rare fluctuations of the potential. 

One possible approach to make the problem numerically accessible is to work in a 3D effective model with a substantially larger unit cell (Ref.~\cite{Heffels2022} uses $1$\,nm-large unit cells). We use instead a different strategy, by modeling the surface of the topological insulator only, thus reducing the numerical complexity considerably, which allows us to treat, e.g., also effects of self-consistent screening.
Such an approach is justified if the fluctuations of the effective disorder potential are smaller than the size of the bulk gap. In insulating bulk samples, the amplitude of these fluctuations is always larger than the gap \cite{Skinner2012,Thomas2017} leading the formations of charged regions (electron-hole puddles) in the bulk. The TI nanowires are, however, so small and the screening by surface states is so efficient, that the potential fluctuations are of the order of $50$\,meV and thus considerably smaller than the bulk gap of $200-300$\,meV \cite{Thomas2017} relevant for Bi based TIs. We will therefore focus on surface states only. 

\begin{figure}[t]
\centering
\includegraphics[width=8cm]{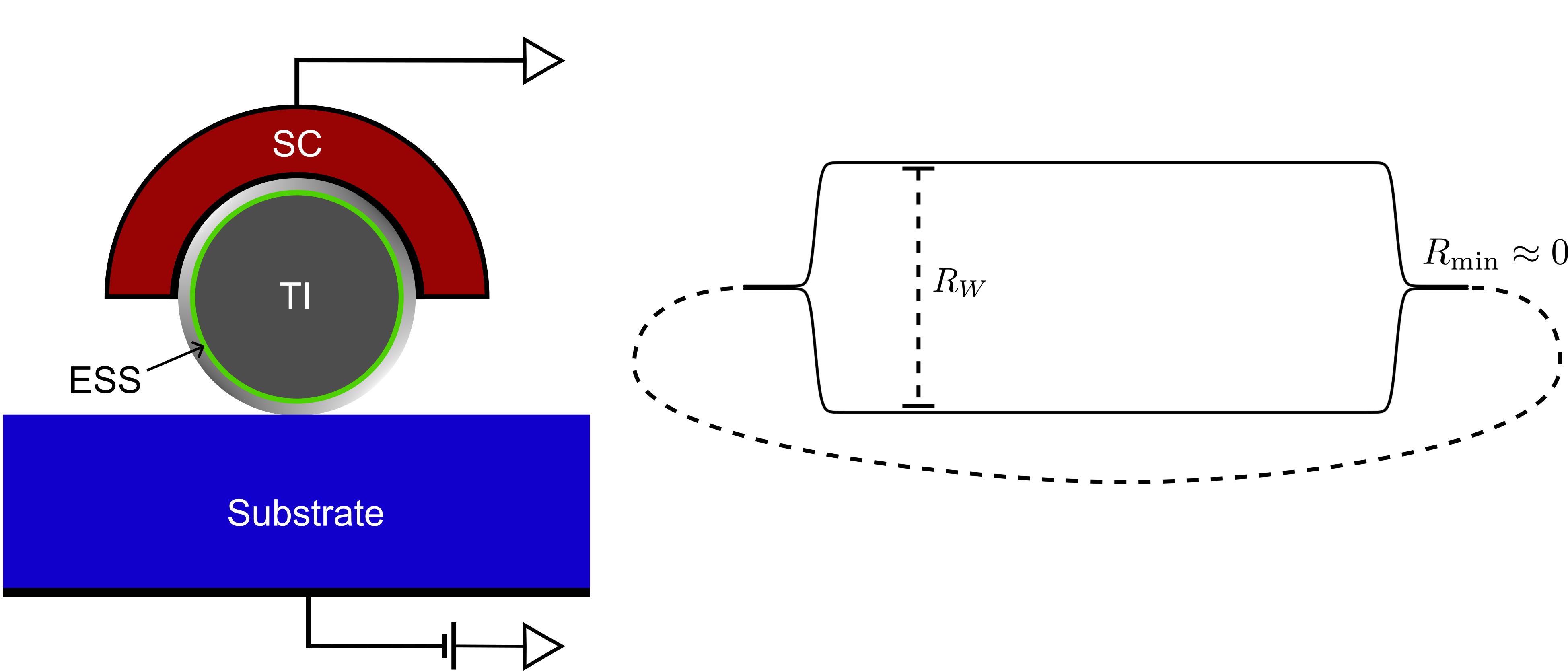}
\caption{Geometrical models of the considered nanowire. Left: z-cut of the wire, illustrating the electrostatic model. The cylindrical TI with radius $R_\text{W}$ is in contact to an insulating substrate below the wire and a half-shell superconductor on top. The gate is modeled via boundary conditions on the bottom side of the substrate and the surface of the SC (see figure). The thickness of the substrate is $1.5\,R_\text{W}$, while the thickness of the superconductor is chosen as $R_\text{W}$. We further simulate a finite penetration depth of the surface states into the wire by evaluating the electrostatic potential at $d\sim5$\AA\ from the surface. For an in-depth discussion of the dielectric environment and the Poisson problem see App.~\ref{AppendixA}. The right panel shows the radial profile used to model the shape of the wire. For simplicity, we assume that the wire radius decays on the scale $R_\text{W}$. The dashed line denotes periodic boundary conditions. The sharp tip at the end of the wire does not affect any physical results as the band gap is large in this region and also the MZM wave function does not penetrate into this region, see Fig. \ref{fig:sampleMajorana}{\bf b}.}
\label{fig:NWSchematic}
\end{figure}
Our goal is to model semi-realistic devices which, importantly, take into account how charges are screened by the dielectric environment, the superconductor and gates. 
Our setting and the dielectric environment is shown in Fig.~\ref{fig:NWSchematic}. Bi-based TI nanowires have a very large dielectric constant, which we model using $\epsilon_\text{W}=200$. The wire is located on top of an insulating substrate 
($\epsilon_\text{S}=4$, typical for SiO$_2$) which separates the wire from a planar bottom-gate, see Fig.~\ref{fig:NWSchematic}. In App.~\ref{AppendixA} we describe how we determine the relevant Coulomb potential of point charges by numerically solving the Poisson equation. 

The surface state of a topological insulator cannot be modeled using a simple real-space tight binding model due to topological constraints. Therefore, we use a description in momentum space, where such a problem is absent. More precisely, we use two different setups, one for screening and one for the analysis of Majorana edge modes. In the first setup, we consider a wire of length $L$ and radius $R_\text{W}$ with periodic boundary conditions where the cylindrical surface state (in the absence of disorder) is described by
\begin{align}\label{eq:Hcyl}
    H=\frac{\hbar v_\text{F}}{R_\text{W}}\left(\sigma_x(l-\varphi)+\sigma_ykR_\text{W}\right).
\end{align}
Here $k=\frac{2 \pi}{L} n$, $n\in \mathbb Z$ is the momentum parallel to the wire, while the angular momentum quantum number $l=m+\frac{1}{2}$,  $m\in \mathbb Z$ takes half-integer values due to antiperiodic boundary conditions arising from the spin Berry phase \cite{Zhang2010}.
The flux through the wire is parametrized in units of the flux quantum, $\varphi=1$ therefore corresponds to one flux quantum $\frac{2 \pi \hbar}{e}$. 
Microscopically, the surface state has a finite depth, decaying exponentially into the bulk of the wire. We approximate this effect by assuming that the surface state has an average distance of $5\,$\AA  \ from the superconductor, consistent with ab-initio results for $\text{Bi}_2\text{Te}_3$ \cite{Ruessmann2022}. This setting is used for our analysis of screening of charged impurities.

The model described above, however, cannot be used to analyze edge modes. Here, we need a model for the ends of the wire. At the same time, we want to keep periodic boundary conditions and a description in momentum space as described above. This problem can effectively be solved by using a wire model in which the radius $R(z)$ is not constant but depends on $z$, the coordinate parallel to the wire, see Fig.~\ref{fig:wire_schematic}. For simplicity, we assume that the wire radius decays exponentially on the length scale set by the radius $R_\text{W}$ of the wire. We use  
\begin{align}
    R(z)=\frac{R_\text{W}-R_\text{min}}{\exp((|z|-z_0)/R_\text{W})+1}+R_\text{min},
\end{align} 
with $R_\text{min}=10^{-2}R_\text{W}$ and periodic boundary conditions, see App.~\ref{Appendix:Curved} for details.
In the region where $R(z)\approx R_\text{min}$, the local band gap $\hbar v_\text{F}/R$ is much larger than any other scale and the wire turns into a trivial band insulator. This effectively models a wire with two ends located at $\pm z_0$ while keeping periodic boundary conditions.

To model this additional curvature along $z$-direction, we have to consider the Dirac Hamiltonian on a curved surface, which is derived in App.~\ref{Appendix:Curved}
and, in position space, is given by
\begin{align}\label{eq:mainCurvedDirac}
    H_D=\hbar v_\text{F}\left(\sigma_x\frac{1}{R}\left(i\partial_\theta-\varphi\right)-\sigma_y\frac{1}{\sqrt{1+R'^2}}\left(i\partial_z+\frac{R'}{2R}\right)\right).
\end{align}
While the Hamiltonian is written here in real space, we transform it to momentum space by Fourier transformation for our numerical implementation as explained above. 

Gating by the voltage $U_\text{G}$ gives rise to an extra potential
\begin{align}\label{eq:gate}
    H_\text{gate}=U_\text{G} u(\theta,z),
\end{align}
which is computed from the Poisson equation, see App.~\ref{AppendixA}, where the potential of the gate is set to $U_\text{G}$ while the potential of the superconductor is set to $0$, matching the Fermi energy of the wire. The latter condition follows from the fact that electrons can tunnel from the wire to the superconductor. As discussed in  Sec.~\ref{sec:unaccounted}, the effect of contact potentials are not included in our study.

The disorder is modeled by placing impurities randomly at positions $\vec r_i$ in the bulk of the wire. 
\begin{align}\label{eq:Hdis}
    H_\text{dis}=\sum_i q_i V(\theta,z, \boldsymbol r_i)
\end{align}
with $q_i=\pm 1$ and the potential $V$ is obtained from solving the Poisson equation for a point charge within the electromagnetic environment of the wire, see App.~\ref{AppendixA}. This, importantly, includes the screening effects by the superconductor, the substrate and the gate.

Furthermore, we study the effects of electron-electron interactions using the Hartree approximation,
\begin{align}\label{eq:Hartree}
    H_\text{Hartree}=\int V(\theta,z,\theta',z') \delta\rho(\theta',z') d\theta'dz', 
\end{align}
where the fluctuation of the electron charge density, $\delta\rho(\theta',z')=\sum_j |\Psi_j(\theta',z')|^2-\rho_0$, is computed self-consistently from the occupied  electronic surface states with wave function $\Psi_j(\theta,z)$.
Here $\rho_0$ is chosen such that $\delta \rho=0$ for the clean system with Fermi level at the Dirac point.
Using the Hartree approximation makes the system computationally tractable. More importantly, the Hartree approximation treats correctly the long-range part of the Coulomb interaction and thus is suitable to study the formation of electron-hole puddles due to charged impurities.

Proximity induced superconductivity is modeled by the s-wave paring amplitude $\Delta(\theta,z)=\Delta_\text{sc}(z) g(\theta)$, where $g$ is 1 below the SC and 0 elsewhere, see Fig.~\ref{fig:NWSchematic} . In  second-quantized form, the superconductivity is described by
\begin{align}
    H_\text{SC}=\sum_{k,k',l,l',\sigma}\Delta_{l,l',k,k'}c^\dagger_{-\sigma,-k',-l'}c^\dagger_{\sigma,k,l}+\text{h.c.},
\end{align}
where \begin{align}\Delta_{l,l',k,k'}=\int d\theta d z \,e^{i (l-l') \theta} e^{i (k-k') z } \Delta(\theta,z)\end{align}
is the Fourier transform of $\Delta(\theta,z)$.

Writing also all other contributions in second-quantized form, the total Hamiltonian of the system is given by 
\begin{align}\label{eq:MZMHamiltonian}
    H_\text{MZM}=H_D+H_\text{gate}+H_\text{dis}+H_\text{Hartree}+H_\text{SC}.
\end{align}
Below, we will first consider the problem without a superconducting term and then focus on the superconductors' impact on the dielectric environment. This will further allow us to show that one can neglect the contribution of $H_\text{Hartree}$ for the geometry studied in our analysis and explore the physics of Majorana modes neglecting the electron-electron interactions.

\section{Potential fluctuations and screening mechanisms}\label{secC}
\subsection{Screening mechanisms}
We first investigate how the impurity potential is screened in our system. As discussed above, we consider here a wire with periodic boundary conditions and constant radius $R$. Furthermore, for the numerical results  of this subsection, we set the gate potential to zero and assume that the Fermi level resides at the bottom of the first band, $\mu_0=\Delta_\text{W}/2$. Changes to either of these parameters only affect the screening by electron-electron interactions, see below.

There are two main sources of screening in our problem. The first channel arises from the surface of the TI and is due to electron-electron interactions. The electronic wave function adapts to the disorder potential, self-consistently forming charge puddles on the surface and consequently reducing the effective potential. This effect is described by the Hartree term, Eq.~\eqref{eq:Hartree}, which is evaluated self-consistently for a given disorder realization. The Coulomb interactions have a second important effect, also described by the Hartree approximation: due to gating, the wire gets charged, thereby renormalizing the gating potential.

The second screening channel is given by the dielectric environment and, most importantly, the superconductor located close to the TI surface. For each charge at distance $d$ to a (planar) superconductor, one obtains a mirror charge at distance $2d$, which leads to a decay of the potential on this scale. This effect is included 
in Eq.~\eqref{eq:Hdis} and \eqref{eq:Hartree} by using the Coulomb potential $V$ computed by solving the Poisson equation with the relevant boundary conditions, see App.~\ref{AppendixA}.

\begin{figure}[t]
\centering
\hspace{-7mm}
\includegraphics[width=8.25cm]{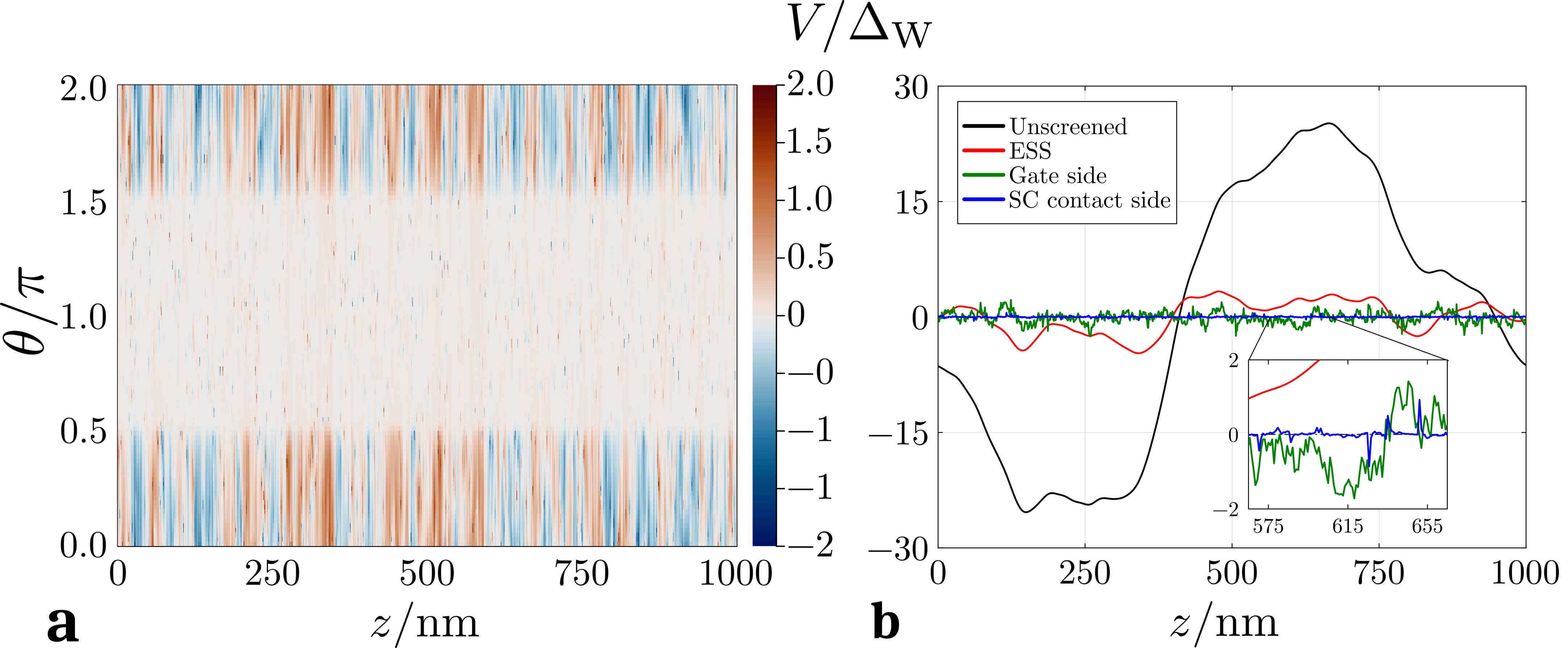}
\caption{Characteristic potential of a nanowire of $R=15$nm radius. \textbf{a}: The full solution of the hybrid wire including metallic boundary conditions and screening by surface electrons.  \textbf{b}: 1D cut along the z-axis of a wire with the same disorder configuration subject to different screening mechanisms. The approximate potential amplitudes range from (in units of $\Delta_\text{W}$) 20 (no screening) over 3 (screening by surface electrons, consistent with \cite{Huang2021}) to 0.5 and 0.1 (superconductor screening, viewed on the gate side and SC contact side, respectively). Note, that even on the side that is in touch with the gate, the superconductor provides screening efficient enough to reduce the disorder potential well below the band gap. The inset in \textbf{b} shows the region in which the potential is strong enough to destroy Majorana modes under certain conditions, see Fig.\ref{fig:sampleMajorana}. Parameters used: $U_\text{G}=0$, $\mu_0=\Delta_\text{W}/2$ and as specified in App.~\ref{AppendixA}.}
\label{fig:samplePotential}
\end{figure}

A typical result for $n_\text{imp}=10^{19}\text{cm}^{-3}$ is shown in Fig.~\ref{fig:samplePotential} where the potential is measured in units of the electronic band gap of the undisturbed wire, $\Delta_\text{W}=\frac{\hbar v_F}{R}$. 
The black curve in Fig.~\ref{fig:samplePotential}{\bf b} shows the bare, unscreened potential. Without screening by electron-electron interactions and without the proximity to the superconductor, the potential fluctuations are gigantic of the order of several tens of $\Delta_\text{W}$. These fluctuations are reduced by an order of magnitude if one considers the screening by electron-electron interactions (red line). This shows that for a  nanowire, which does {\em not} have a superconductor close by, screening by electron-electron interactions is extremely important. The resulting effective disorder potential is, however, still sizable. The amplitude of potential fluctuations is $\Gamma=\sqrt{\langle V^2\rangle}\sim 2.5 \Delta_\text{W}$ in this case (consistent with analytical estimates of Ref.~\cite{Huang2021}). Naively, this suggests that stable Majorana zero modes cannot be realized in such a system.

The situation does, however, change drastically when also the screening effect due to the proximity of the superconductor are taken into account. Even on the side opposite to the superconductor (green line), potential fluctuations are reduced by another factor of $5$, $\Gamma \sim 0.5\, \Delta_\text{W}$.
More dramatically, directly below the superconductor, potential fluctuations are almost absent, $\Gamma\sim 0.05\, \Delta_\text{W}$. This is also shown in Fig.~\ref{fig:samplePotential}{\bf a}, which shows the disorder potential for one impurity realization as a function of $z$ and the azimuthal angle $\theta$. Below the superconductor, $0.5 \pi < \theta < 1.5 \pi$, there are only tiny fluctuations which become much larger outside of this range. The  screened potential fluctuates on a length scale set by the wire radius, see Fig.~\ref{fig:AppendixScreeningComparison} in App.~\ref{AppendixA2}.
The order-of-magnitude difference of the amplitude of the disorder potential on the different sides of the wire is an important effect which has to our knowledge not been considered before and which is absent in more phenomenological models of disorder~\cite{Heffels2022}.

The presence of a nearby superconductor therefore has two main consequences: first and most importantly, it reduces the potential fluctuations well below the band gap, even on the side of the TI farthest away from the SC. This can already be taken as an indicator, that band physics (such as the formation of Majorana states) may be preserved for the chosen impurity density and wire radius. Secondly, the modification to Coulomb potential also reduces drastically the interactions between the surface electrons in the same manner and, by extension, also the screening provided by them. 

Consequently, the question arises, whether the screening channel by electron-electron interaction is relevant for the device geometry considered by us. The answer is that electron-electron screening has only negligibly small effects for our device. This is shown in Fig.~\ref{fig:samplePotential} and further analyzed in App.~\ref{AppendixA2}. In the presence of the superconductor, we find that corrections from electron-electron interactions, which are gigantic in the absence of the superconductor, become smaller than $5\%$ and thus negligible in the presence of the superconductor. This arises from a combination of three effects: \textit{a)} the electron-electron interactions are screened by the proximity to the superconductor, \textit{b)} due to SC screening the amplitude of potential fluctuations is relatively small, and \textit{c)} the potential fluctuates on a length scale set by the distance to the SC, $\sim R_\text{W}$, while relevant Fermi wavelengths are typically a factor of 5-10 larger. Thus, the electronic wavefunction in this system simply cannot locally adjust to the disorder potential. 
One can quantify this effect by considering the dimensionless ratio 
\begin{align} 
\frac{V_{ee}(k=2\pi/R)}{\hbar v_\text{F}}\approx 0.007\ll 1,
\end{align} 
where $V_{ee}(k=2\pi/R)$ is the Fourier transform of the effective Coulomb potential on the gate side. The main reason for the smallness of this parameter is the large value of the dielectric constant ($\epsilon\approx 200$) in the topological insulator considered by us.

\begin{figure}[t]
\centering
\includegraphics[width=8.5cm]{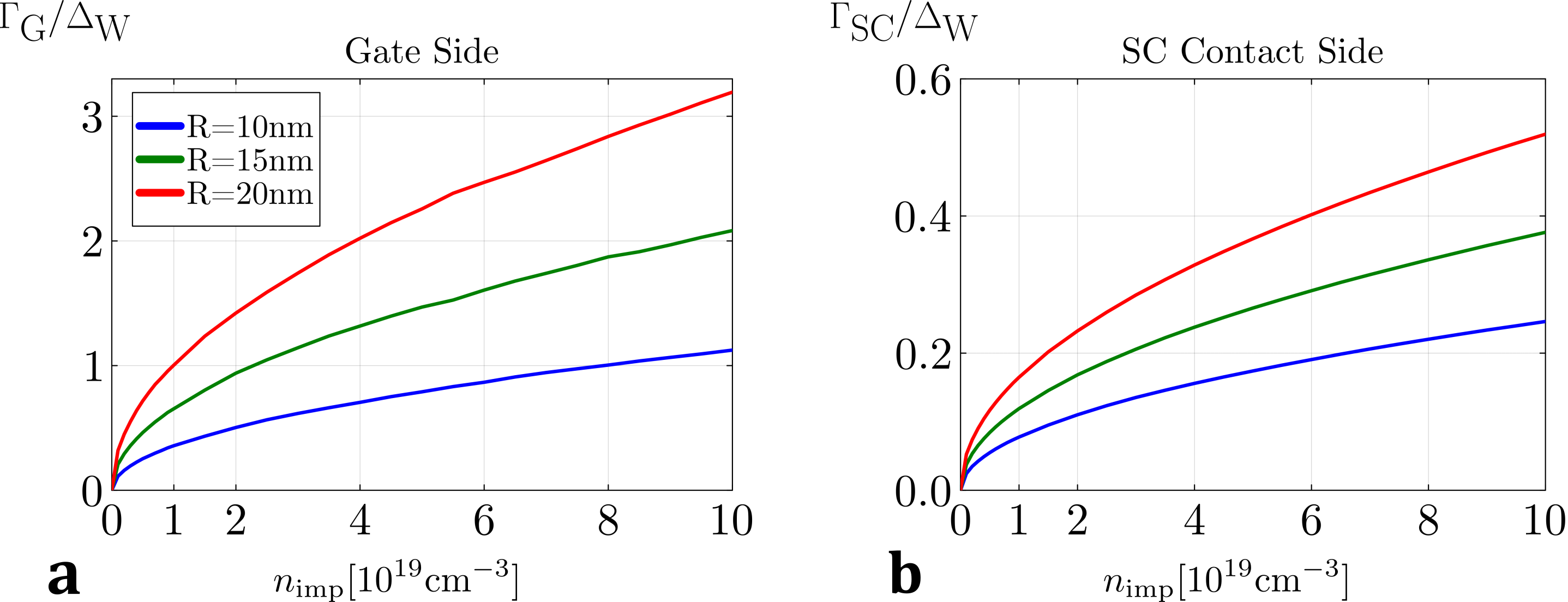}
\caption{Sample-averaged potential amplitudes as functions of impurity density for wires of radii $R=10,15,20\,$nm. While the typical potential quickly exceeds the band gap on the gate side, it remains much smaller on the SC contact side, even for large impurity densities. Additionally, we find, that, on the gate side, there is still an equivalence between \textit{i)} multiplying the radius by a factor $a$ and \textit{ii)} multiplying the impurity density by a factor $a^3$ (see main text).}\label{fig:V_vs_n}
\end{figure}

\subsection{Potential fluctuations}
As screening by surface electrons can be neglected in the presence of a superconductor, one can calculate the average amplitude of potential fluctuations analytically using
\begin{align}\label{eq:fluctAna}
\langle V^2(\vec r)\rangle&=\left\langle \left(\sum q_i V_s(\vec r,\vec r_i)\right)^2\right\rangle \nonumber \\
    &= n_\text{imp} \int_\text{wire} (V_s(\vec r,\vec r'))^2 d^3 \vec r',
\end{align}
where $\langle \dots \rangle$ denotes the average over impurity positions $\vec r_i$ and impurity charges $q_i$
with $\langle q_i q_j \rangle=\delta_{ij}$, $V_s(r,r_i)$
is the impurity potential screened by the electromagnetic environment. This is, in general, a complicated function which we can be calculated only numerically. Close to the surface of the superconductor the potential becomes short-ranged and we can approximate the curved surface by a planar superconductor. In distance $d$ from the superconductor, the potential of a single impurity is then given by 
\begin{align}
    V_s(d) \approx \frac{1}{4 \pi \epsilon_0 \epsilon}\bigl(\frac{1}{(r_\text{imp}^2-(d-d_\text{imp})^2)^{1/2}}\nonumber \\-\frac{1}{(r_\text{imp}^2-(d+d_\text{imp})^2)^{1/2}}\bigr).
\end{align}
Using this result and performing the integral in Eq.~\eqref{eq:fluctAna}, we obtain close to the superconductor
\begin{align}\label{eq:GammaSC}
  \Gamma|_\text{SC side}=  \sqrt{\langle V^2(d) \rangle}&\approx \frac{e^2\sqrt{4 \pi}}{4 \pi \epsilon_0 \epsilon} (n_\text{imp} d)^{1/2}  \\
    &\approx 1.8\,\text{meV} \left(\frac{n_\text{imp}}{10^{19} \text{cm}^{-3}}\right)^{1/2} \nonumber\\
    &\approx 0.068\, \Delta_\text{W} \frac{R}{10\,\text{nm}} \left(\frac{n_\text{imp}}{10^{19} \text{cm}^{-3}}\right)^{1/2}\!\!,\nonumber
\end{align}
 where we used $d\approx 5$\AA. In the last line, we have written the potential fluctuations in units of the (surface) band gap $\Delta_\text{W}$ of a TI nanowire of radius $R$. The analytic result is in quantitative agreement with our numerical results shown in Fig.~\ref  {fig:V_vs_n} with deviations smaller than $10\,\%$ arising mainly from the curvature of the surface. The plot and the analytic formula show that directly
 below the superconductor potential fluctuations remain small  for realistic impurity densities of the order of $10^{19}$\,cm$^{-3}$ and wire radii of several tens of nanometers.

The potential fluctuations are, however, much bigger on the gate side, opposite to the superconductor. In this case, $d$ is not important anymore and thus $R$ is the only relevant length scale. Therefore,  $\Gamma/\Delta_\text{W}$ can only depend on the dimensionless impurity density $R^3 n_\text{imp}$. We obtain
\begin{align}\label{eq:Gamma_gate}
  \Gamma|_\text{gate side}
    &\approx  0.13\, \Delta_\text{W}  \left( R^3 n_\text{imp}\right)^{1/2}
    \\ &\approx  0.41\, \Delta_\text{W} \left(\frac{R}{10\,\text{nm}}\right)^{3/2} \left(\frac{n_\text{imp}}{10^{19} \text{cm}^{-3}}\right)^{1/2},\nonumber
\end{align}
where the prefactor was not calculated analytically, but fitted to the numerical result shown in Fig.~\ref{fig:V_vs_n}. The prefactor is expected to depend on details of the geometry of the system and also on the value of the dielectric constant of the TI. For our parameters, the disorder potential on the gate side is thus roughly a factor 5-7 larger compared to regions close to the superconductor. A similiar estimate has been obtained analytically in Ref.~\cite{Huang2021}.

Which of the disorder potentials is most relevant for the stability of Majorana fermions? This depends on the location of the Majorana wave function which is strongly influenced by gating which we therefore consider next.

\begin{figure}[t]
\centering
\includegraphics[width=8.5cm]{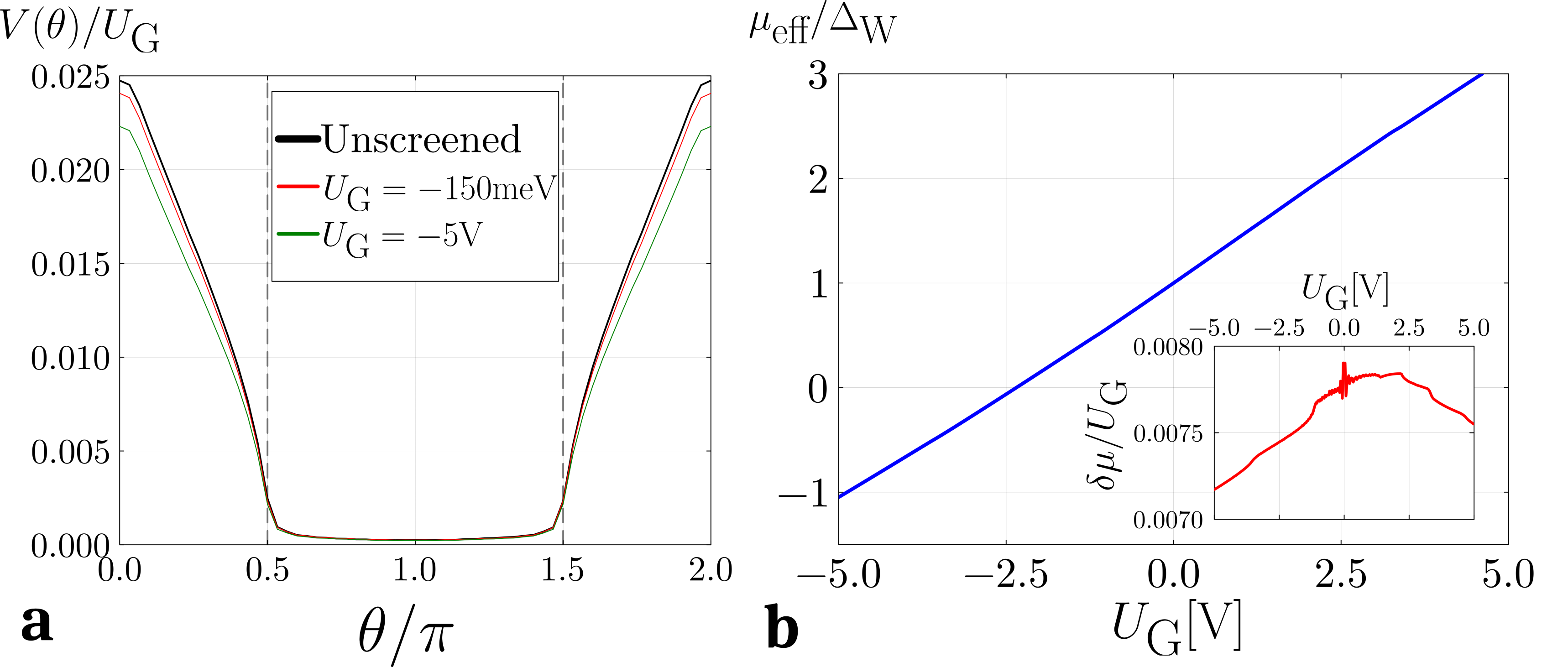}
\caption{Panel {\bf a}: Gate induced potential as function of the azimuthal angle $\theta$ in units of the applied gate voltage $U_\text{G}$ for a wire with $R=15\,$nm. $V=0$ corresponds to the potential of the superconductor. Directly below the superconductor, $0.5 < \theta/\pi < 1.5$ the potential remains close to zero but grows to $0.025 \, U_\text{G}$ on the opposite site for the geometry shown in Fig.~\ref{fig:NWSchematic}. The strong reduction in $V(\theta=0)$ arises mainly due to the large dielectric constant, $\epsilon=200$, of the TI. The black curve shows the potential obtained when ignoring the screening due to electrons in the TI surface states. In this case $V(\theta)/U_\text{G}$ is independent of $U_\text{G}$. When screening due to the TI surface state is taken into account (red and green curve for $U_\text{G}=-150\,$meV and $-5\,$V, respectively, for a wire with a radius of $15$\,nm), the potential is slightly reduced. The screening is larger for larger voltages as in this case more electrons contribute to screening. Panel {\bf b}: Average potential $\mu_\text{eff}$, Eq.~\eqref{mueff}, in units of TI-surface band gap $\Delta_\text{W}$.  The inset shows for $R=15\,$nm, that due to screening the ratio $(\mu_\text{eff}-\mu_0)/U_\text{G}$ depends weakly on $U_\text{G}$. Kinks in the curve arise from the (de-)activation of surface bands.}\label{fig:gate_plots}
\end{figure}

\subsection{Gate potentials}\label{sec:gate_potential}

In an experimental system, the position of the Fermi level in the absence of an external gate voltage,  $U_\text{G}=0$, depends on the precise doping of the wire, for definiteness we set the chemical potential to $\mu_0=\Delta_\text{W}$ for $U_\text{G}=0$. This parameter allows to create stable topological superconductivity without too much gating, which in turn allows to keep the cutoffs in our calculation small. Additionally, this estimate is not too far off the initial Fermi level in experimental wires \cite{Muenning2021}, which is typically only a few band gaps away from the Dirac point.
By changing the gate voltage $U_\text{G}$, we can manipulate the position of the effective chemical potential in the wire and, importantly, modulate the band structure.

In Fig.~\ref{fig:gate_plots}{\bf a} we show the resulting gate potential as a function of the azimuthal angle $\theta$. The potential is small and approximately constant below the superconductor, $\pi/2 <\theta<3 \pi/2$ but grows rapidly on the opposite site. The figure also shows that the screening of the gate potential from the TI-surface states (calculated within a self-consistent Hartree approximation for a clean system) is a small effect. This can again be traced back to the presence of the SC, more precisely the renormalization of the electron-electron interactions. 

The phase diagram is largely controlled by the effective chemical potential, see Fig.~\ref{fig:gate_plots}{\bf b},
\begin{align}\label{mueff}
    \mu_\text{eff}=\mu_0 + e  \int V(\theta)d\theta/(2 \pi)
\end{align}
computed from the average potential, where $e$ is the elementary charge (defined positive). 
Neglecting the screening effects, we find for our geometry
\begin{align}
    \mu_\text{eff}\approx \mu_0+0.0075\,e U_\text{G},
\end{align}
independent of wire radius.
Corrections to this formula are small and even in the presence of screening there is a roughly linear relation between the applied gate voltage and $\mu_\text{eff}$, see Fig.~\ref{fig:gate_plots}{\bf b}.
The inset shows deviations from this linear behavior, arising from the electronic screening.  The effect gives a correction of roughly $10\,$\%, and we use the fully screened gate potential in the following calculations (at $\varphi=0$), while the much smaller screening effects of the disorder potentials are neglected.

\section{Stability of Majorana modes}\label{secD}

\subsection{Phase diagram of the clean system}\label{sec:clean}
To obtain topological superconductivity and the associated Majorana zero modes, 
we now consider the proximity effect to the superconductor and the impact of an external magnetic flux oriented parallel to the wire as discussed in Sec.~\ref{secB}. It has recently suggested by Legg {\it et al.} \cite{Legg2021}  that to obtain stable Majorana modes at small magnetic fields, one should also make use of gate-induced electric fields which leads to an extra Rashba-like splitting of surface bands. In our model, we calculate the relevant potential (which was introduced as a phenomenological parameter in Ref.~\cite{Legg2021}) microscopically, see Sec.~\ref{sec:gate_potential}. Changing the gate voltage $U_\text{G}$ changes both the average chemical potential and also inhomogeneous electric fields inside the wire, see  Fig.~\ref{fig:gate_plots}. 

For our simulations, we fix the induced superconducting pairing potential to be $\Delta_\text{SC}=1.6\,\text{meV}$, consistent with the proximity effect in $\text{Bi}_2\text{Se}_3$ and Nb \cite{Floetotto2018}. 

\begin{figure}[t]
\centering
\includegraphics[width=8.25cm]{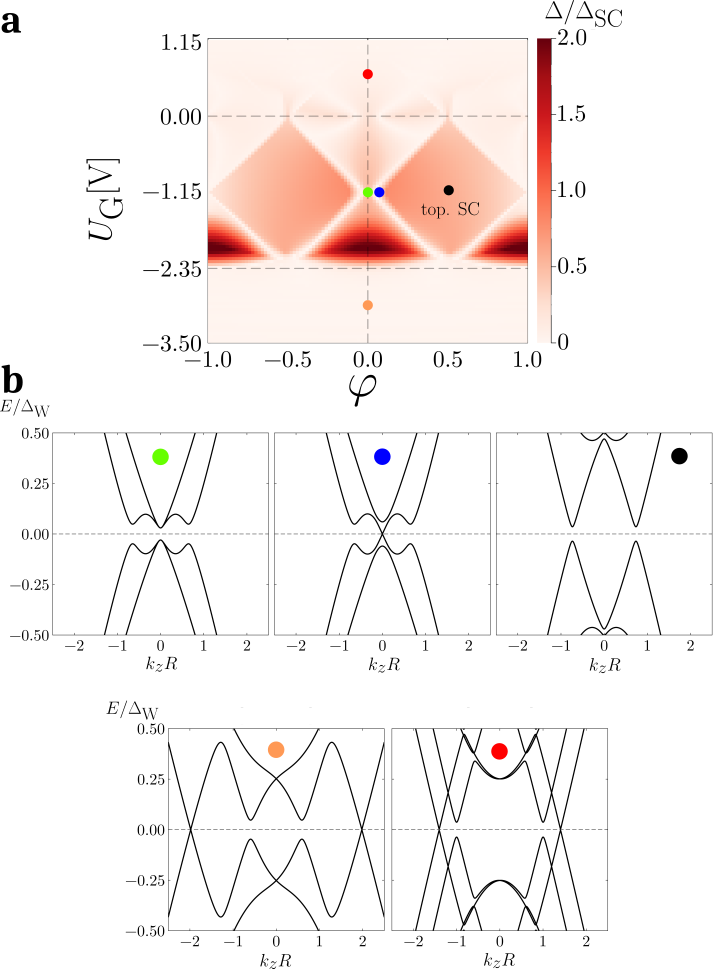}
\caption{Bulk gap of a disorder free proximitized nanowire as function of flux $\varphi$ and gate voltage $U_\text{G}$.
Only for $0\gtrsim U_\text{G} \gtrsim -2.35$\,V the electronic low-energy states resides below the superconductor, thus obtaining a sizable superconducting gap, see main text. Panel {\bf b} shows the band structures at selected points in the phase diagram as indicated by colored dots.
}\label{fig:phaseDiagramClean}
\end{figure}
We can now discuss the phase diagram and the band structures of the clean system. In Fig.~\ref{fig:phaseDiagramClean}{\bf a}, we show the gap of the proximitized system as function of gate voltage $U_\text{G}$ and flux $\varphi$ for a wire of radius $R=15$\,nm. For vanishing gate voltage, we assume that the chemical potential is located in the first valence band of the TI surface state, $\mu_0=\Delta_\text{W}$. 

We first observe that both for $U_\text{G}>0$ and for $U_\text{G}<-2.35\,$V, the gaps become very small. This can be explained by the fact that a positive gate voltage pushes electrons towards the gate, away from the superconductor. Therefore, for  $U_\text{G}>0$, electrons close to the Fermi surface weakly hybridize with the superconductor. For  $U_\text{G}<-2.35\,$V the effective chemical potential $\mu_\text{eff}$ becomes negative, see Fig.~\ref{fig:gate_plots}, and charge carriers at the Fermi surface are holes rather than electrons, which are attracted by the positively charged gate. Therefore, again, states at the Fermi surface hybridize only weakly with the superconductor. 
The arguments given above, do not apply to the $k_z=0$ states, which always have a constant density, $|\Psi_{k_z=0}(\theta)|^2=\text{const}.$, as a function of the azimuthal angle $\theta$ due to the Klein-tunneling effect \cite{Ziegler2018}. As shown in the last two panels of Fig.~\ref{fig:gate_plots} the strong suppression of the superconducting gap occurs at large momenta, where the Klein-tunneling argument does not apply.
For   $0\gtrsim U_\text{G} \gtrsim -2.35\,$V, the relevant low-energy states reside directly below the superconductor and thus the proximity effect is strongest. In this regime, we observe a pronounced topological phase inside the two clearly visible `diamonds'. At the white lines encircling this region, the topological gap closes, see second plot in Fig.~\ref{fig:phaseDiagramClean}{\bf b} marked by a blue dot. Excluding the region very close to the topological phase transition, the topological gap inside the diamond is sizable, reaching values of up to 80\% of the superconducting pairing potential $\Delta_\text{SC}$ induced by the proximity effect. 
The size of the topological gap is determined by modes with a momentum of order $1/R$, see Fig.~\ref{fig:phaseDiagramClean}{\bf d}, while the gap closing occurs at $k=0$, as shown in Fig.~\ref{fig:phaseDiagramClean}{\bf c}.
Note that our phase diagram and the resulting band structures differ qualitatively from similar pictures in the literature \cite{Heffels2022,Legg2021}
which discuss the effects of chemical potentials. In our model the gate voltage not only shifts $\mu_\text{eff}$ but also induces potential gradients thus inducing Rashba coupling and pushes electrons and holes either towards to away from the superconductor, an effect also discussed in Ref.~\cite{LeggMetallization2022}.

Our analysis shows that to reach stable topological superconductivity in TI nanowires partially covered by a superconductor, it is important to choose the sign of the gate voltage such that all relevant low energy modes -- and therefore also the Majorana zero mode -- reside directly below the superconductor to guarantee a good proximity effect. This has the added benefit that in this region also disorder effects are strongly suppressed by screening as discussed above.

\subsection{Topological superconductivity in the presence of disorder}\label{sec:topoDis}

\begin{figure}[t]
\centering
\includegraphics[width=8.5cm]{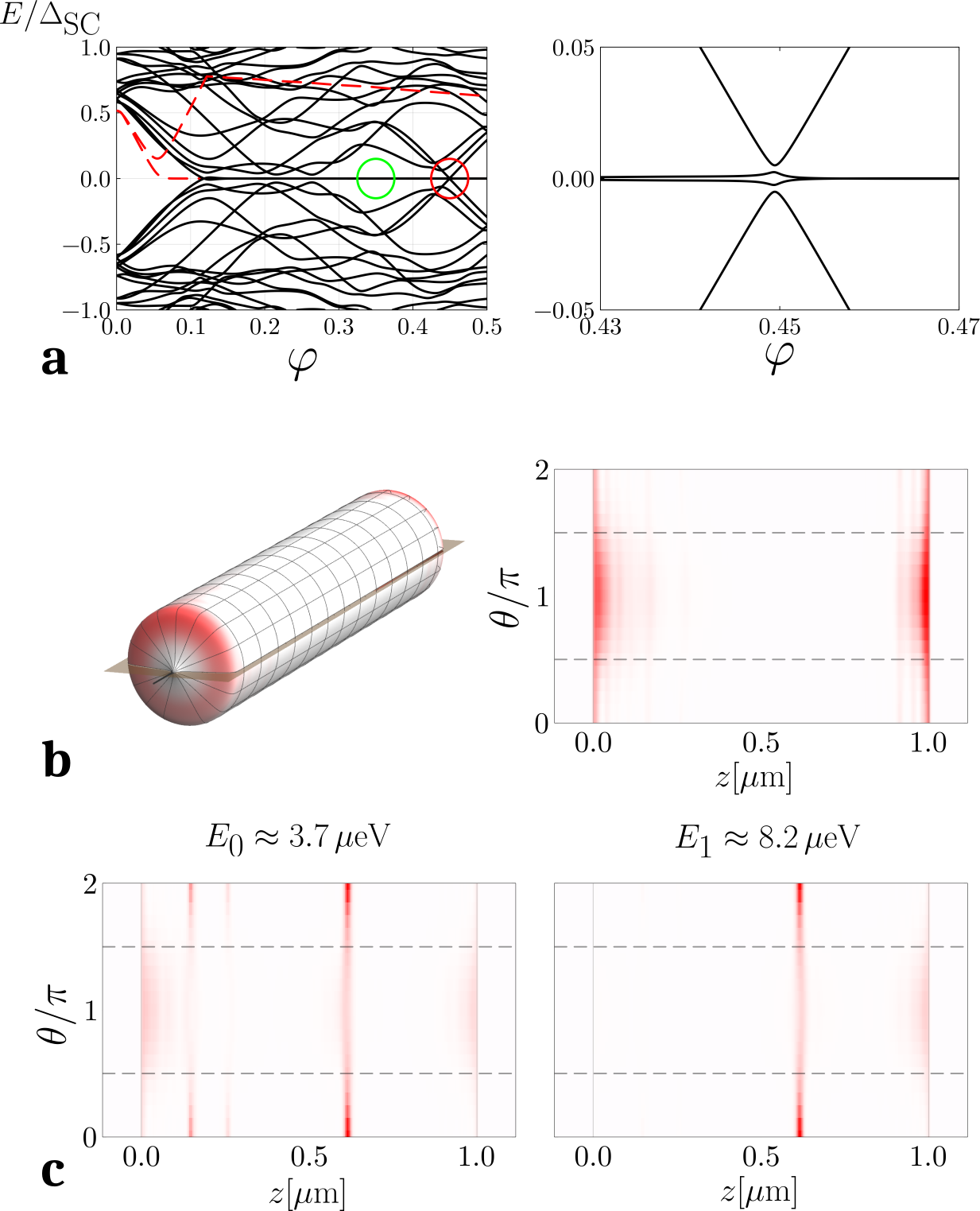}
\caption{Spectra and wave functions of low-energy states for a fixed disorder configuration (same potential as in Fig.~\ref{fig:samplePotential}, $R=15\,$nm, $U_\text{G}=-1.15$\,V, $n_\text{imp}=10^{19}\, \text{cm}^{-3}$). Panel \textbf{a}: Bogoliubov spectrum as function of the magnetic flux.  Due to the disorder, bound states emerge within the gapped region of the clean system (red dashed lines: two smallest energies in a clean wire).
The green circle shows a region with robust Majorana zero modes. The red circle marks a region, where an in-gap state approximately crosses zero energy. Right panel: enlarged view of this region, showing the hybridization of the Majorana zero mode with the excited state.
 Panel \textbf{b}: Real space distribution of the probability density of the Majorana zero mode obtained for $\varphi=0.35$ (green circle in panel {\bf a}). The Majorana mode is mainly located at the end of the wire.
 Right panel: probability distribution as a function of $z$ and the 
 azimuthal angle $\theta$. The wavefunctions has a higher weight directly below the superconductor ($\pi/2<\theta<3 \pi/2$) but extends also to the other side of the nanowire.
 Panel \textbf{c}: Wave functions of the two   lowest-energy states at $\varphi\approx0.45$ 
 ($E_0\approx 3.7\,\mu\text{eV}\approx 0.0023\,\Delta_{\text{SC}}$, $E_1\approx 8.2\,\mu\text{eV}\approx 0.0051 \,\Delta_{\text{SC}}$), 
 where the Majorana state hybridizes with a defect state localized around $z\approx615$\,nm. The inset of Fig.~\ref{fig:samplePotential}{\bf b} shows at this location a strong attractive potential on the gate site of the wire. This is an example of a parameter regime, where no Majorana zero mode can be identified.\label{fig:sampleMajorana}}
\end{figure}

Next, we consider the effects of charged impurities at a realistic disorder density of $10^{19}\, \text{cm}^{-3}$.
In Fig.~\ref{fig:sampleMajorana}, we analyze
a disordered wire with a radius of $R=$15\,nm and a length of $1\,\mu$m. The value  of the gate voltage is fixed to $U_\text{G}=-1.15\,$V, which corresponds to $\mu_\text{eff}\approx 0.5\,\Delta_\text{W}$, the optimal value for topological superconductivity in the clean system, see Fig.~\ref{fig:phaseDiagramClean}.

In Fig.~\ref{fig:sampleMajorana}{\bf a}, we show the Bogoliubov spectrum as a function of the flux $\varphi$ (in units of the flux quantum) for one fixed disorder configuration with roughly 7000\, charged impurities (same as in Fig.~\ref{fig:samplePotential}).
For this disorder configuration, an approximate zero mode emerges for $\varphi\gtrsim 0.12$. The red dashed line marks the size of the topological bulk gap of the clean system. In addition to the state at zero energy, we observe in the presence of disorder only a small number of states below this gap. In the shown example, there are only 5 to 10 such states in a wire with $10^4$ charged defects. The ratio of wire length and correlation length of the disorder (on the gate-side) is of the order of $L/\xi=100$, see App.~\ref{AppendixA2}, but only rare disorder fluctuations create in-gap states.

The green circle marks a regime with a well pronounced gap to excited state, indicating robust Majorana zero modes.
Indeed, a plot of the Majorana wave-function,  Fig.~\ref{fig:sampleMajorana}{\bf b}, shows that the Majorana is well localized at the edges of the sample or rather as rings on the `lids' of the cylinder (see the 3d plot), located  at $z=0$ and 1\,$\mu$m, respectively. This is an example of a `robust' Majorana mode, defined more precisely below.

In contrast, the red circle indicates a region where an in-gap state appears to cross zero energy. An enlarged blow-up of this region, Fig.~\ref{fig:sampleMajorana}{\bf b}, shows an avoided crossing and, simultaneously, a small splitting of the Majorana zero mode. This arises because the edge Majorana modes both have a weak hybridization with the in-gap  state.
This is confirmed when we plot the wave function of the two states with the lowest energy at the location of the avoided crossing, $\varphi\approx0.45$, in  Fig.~\ref{fig:sampleMajorana}{\bf c}.
The defect state, approximately located at $z=615\,$nm, hybridizes most strongly with the right Majorana mode (right panel) forming the state with higher energy, $E_1\approx8.2\,\mu$eV. The low-energy state, $E_0\approx3.7\,\mu$eV, (left panel) is formed by the left Majorana and the defect state. This is an example of a Majorana mode, which cannot be used in a quantum device for two reasons: first, it shows a small splitting of energy and, second, the low-energy Majorana is not located at the edges of the sample anymore.

As discussed in Sec.~\ref{sec:clean},  for the chosen sign of the gate voltage, the wave function of the Majorana mode has a substantial weight below the superconductor, $\pi/2
<\theta< 3 \pi/2$, inside of the white dashed lines. The shown defect state of Fig.~\ref{fig:sampleMajorana}{\bf c} originates from an accidentally strong potential fluctuation on the gate-side of the device.
We will argue below based on an extended statistical analysis that the disorder on the gate-side of the wire is mainly responsible for the degradation of Majorana zero modes.

\subsection{Searching for robust Majorana modes}\label{sec:robust}
The main goal of our paper is to explore whether in the presence of a realistic level of charged impurities, one can realize Majorana modes which can serve as building blocks of topological quantum devices.

Thus, we need to define a set of criteria under which a pair of Majorana zero mode exists in a way that it is (potentially) usable in quantum devices. We label those states as `robust' Majorana zero modes. As we want to perform a statistical analysis, we need a well-defined algorithmic definition of `robust'. 
While the criteria defined below are to some extent arbitrary, we strive for a relatively conservative approach, focusing on states with clearly defined Majorana edge modes.

A main challenge is, that approximate zero energy states can have many sources and  occur frequently even in ordinary, non-topological superconductors \cite{Bagrets2012,Liu2012,Sau2013,Kells2012,Neven2013,Prada2020,Hess2023,Pan2022,Yu2021}. Clearly, robust Majorana zero modes should be localized close to the two ends of the wire. Furthermore, the gap to all other excitations should be as large as possible as quasiparticle poisoning by thermal excitations can be a major source of dephasing, especially when inelastic processes, e.g. due to phonon scattering, are considered.

Taking all this into account, we use the following conditions to distill a simple  answer to the question of whether we can clearly identify in our numerics well-protected Majorana edge modes. In the following, we will denote the states defined below as `robust' MZMs.

(i) First, we use a relatively strict condition for allowed values of the energy splitting of the Majorana modes. We demand that the energy splitting of the Majorana mode remains small, \begin{align}\label{eq:MZM_criteria}
    |\Delta_\text{M}| < 4\cdot10^{-3} \Delta_\text{SC} \approx 6.4\,\mu \text{eV}.
\end{align}
This corresponds to a frequency of $1.6\,$GHz. To put this value into context, for a clean system in the range between $\varphi=0.25$ and $\varphi=0.5$ flux quanta, we obtain for our parameters $|\Delta_\text{M}|\approx 10^{-4}\Delta_\text{SC}$. If  smaller values of $\Delta_\text{M}$ are needed, one needs to use longer wires.

\begin{figure}[t]
\centering
\includegraphics[width=8cm]{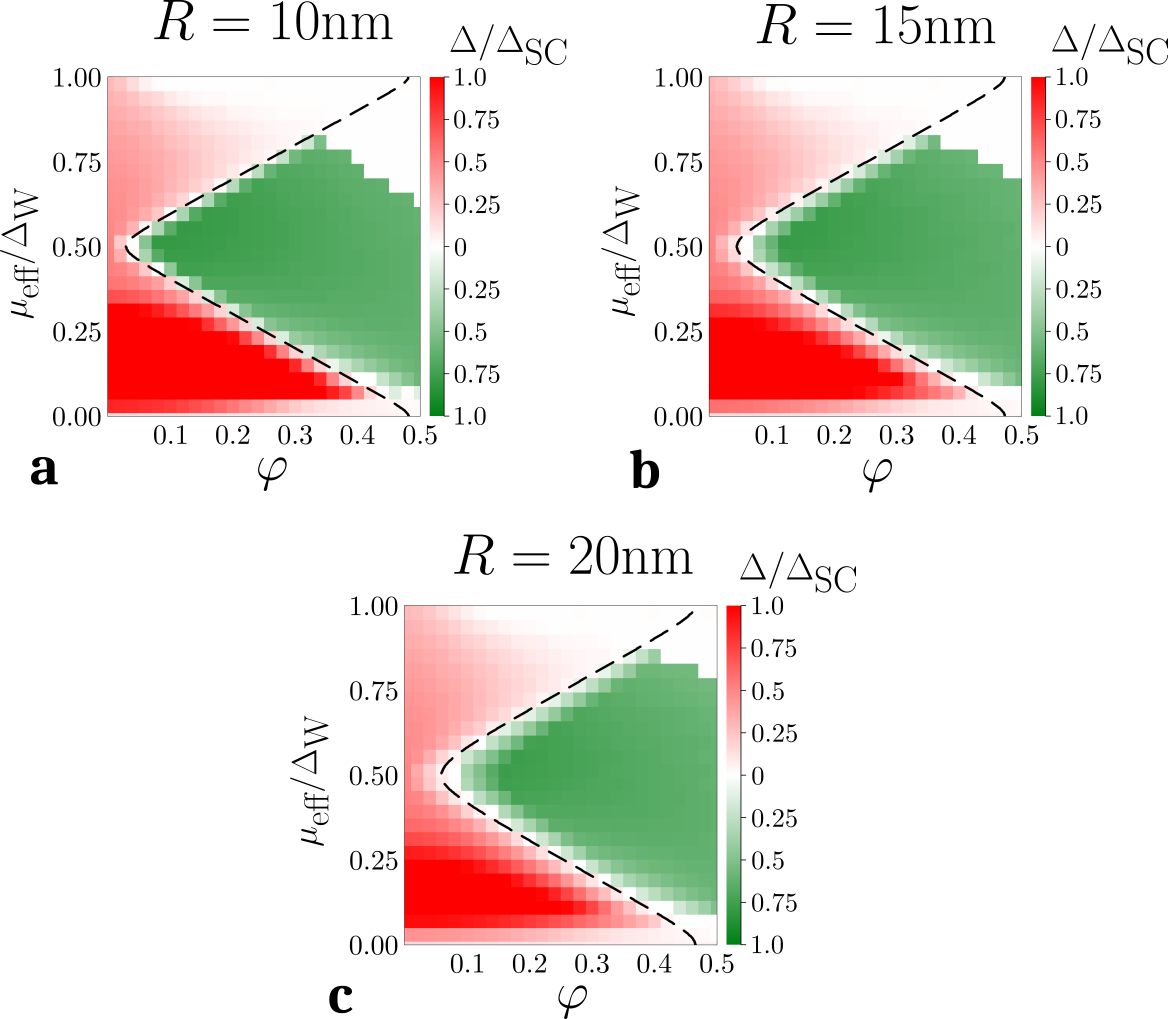}
\caption{\label{fig:GapMapClean}Phase diagrams of three clean wires with different radii as a function of the flux $\varphi$ (in units of the flux quantum) and of the effective chemical potential $\mu_\text{eff}$ controlled by the gate voltage, see Fig.~\ref{fig:gate_plots}{\bf b}.
In cases where we detect a robust Majorana zero mode according to the criteria (i-iii), we show in green shades the size of the topological gap, i.e., the energy difference $E_2-E_1$ between the next excited state $E_2$ and the Majorana energy $E_1\approx 0$. Here we denote by $E_n$ the $n$th positive energy in the Bogoliubov spectrum of the wire. In all other cases, white to red colors encode the value of the (superconducting) gap $E_1$. The black dashed lines indicates the phase transition in the bulk system. The white region in the top right corner of the three figures arises because the wave function is not sufficiently localized close to the ends of the wire, see main text. Parameters: as in Fig. \ref{fig:sampleMajorana}.} 
\end{figure}

(ii) The topological gap $\Delta$ between the state at approximately zero energy and the first state above should  be sufficiently large,
\begin{align}
    \Delta >0.05\,\text{meV} \approx 0.03 \Delta_\text{SC} \approx k_\text{B}\cdot 0.5\,\text{K}. 
\end{align}
For these value, the thermal occupation of the excited state at an effective temperature of 50\,mK is smaller than $e^{-0.5/0.05}\approx 5\cdot 10^{-5}$, while it is below $10^{-7}$ at 30\,mK.

(iii) Furthermore, the Majorana zero mode should be localized {\em both} close to the left and the right edge of the wire. We demand that
\begin{align}\label{eq:MZMcriterion3}
\int_0^{2 \pi} d\theta   \int_{-b/2}^\xi dz\,|\Psi_\text{M}(z,\theta)|^2  &>0.3, \ \text{and, simultaneously,}\nonumber\\
\int_0^{2 \pi} d\theta   \int_{L-\xi}^{L+b/2} dz\,|\Psi_\text{M}(z,\theta)|^2  &>0.3, \quad \xi=50\,\text{nm,}
\end{align}
where $b/2\approx0.125L$ is added to the left and right end of the wire, see App.~\ref{Appendix:Curved}. $\Psi_M(z,\theta)$ is the normalized eigenfunction obtained from the diagonalization in Bogoliubov space (which therefore includes both particle and hole components).  
In the clean system, the value of the integrals is between $0.35$ and $0.4$ for fluxes in the range of $0.25 \lesssim \varphi \lesssim 0.5$ at the optimal gate voltage $\mu_\text{eff}/\Delta_\text{W} \approx 0.5$.

The set of conditions defined above allows for a clear identification of topological Majorana modes. Conditions (i) and (ii) guarantee that there is a Majorana zero modes well separated from the rest of the spectrum and condition (iii) ensures that the two Majorana operators $\gamma_L$ and $\gamma_R$ are located on the left and right side of the wire. To be able to contact the Majorana modes experimentally, it is also important that they are located at the edges of the wire.

Another possible criterion to identify topological phases is to compute the topological invariant characterizing topological superconductors \cite{Kitaev2001}. This involves the computation of a Pfaffian, which is numerically much more expensive than the diagonalization of the system. In practice, our Bogoliubov matrices are of size
 $10000 \times 10000$, which makes the computation of the topological invariant too expensive. We have, however, computed the Pfaffian topological invariant for a system with 
a different set of boundary conditions, see App.~\ref{app:pfaffian}, which allows to use a smaller cutoff in $k$-space leading to smaller matrices. These calculations confirm the validity of the criteria above. 
The criteria given above turn out to be much stricter than the identification of the topological phase by the Pfaffian as shown in the Appendix. In almost all cases where we found a `robust Majorana' according to the criteria above, this could be confirmed by the Pfaffian, see App.~\ref{app:pfaffian}.

\begin{figure}[t]
\centering
\includegraphics[width=8cm]{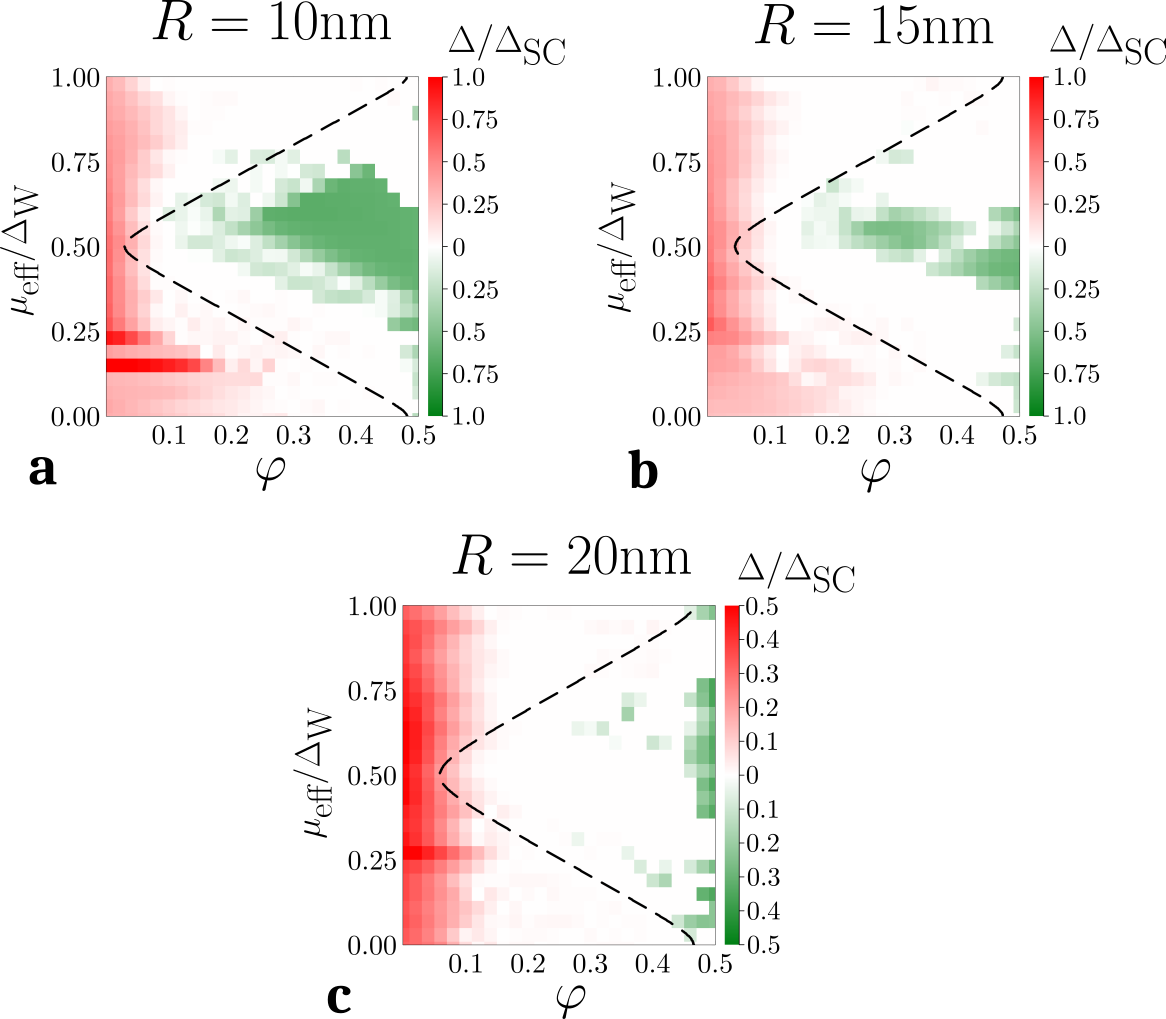}
\caption{\label{fig:GapMap}Phase diagrams of three disordered wires ($n_\text{imp}=10^{19}\,\text{cm}^{-3}$) with different radii (color coding, inset lines and parameters analogous to Fig.~\ref{fig:GapMapClean}). The most stable regions reside around $\mu_\text{eff}=\Delta_\text{W}/2$, where the tolerance for the variation in the local chemical potential is largest. Note that panel {\bf c} uses a different color scale than panels {\bf a} and {\bf b} for better visibility.} 

\end{figure}

With this recipe we study the emergence of robust Majorana modes as a function of gate voltage and magnetic flux followed by a statistical analysis of our results.

We first study the emergence of Majorana modes in a perfectly clean system,  Fig.~\ref{fig:GapMapClean}. Green shades are used whenever our algorithm identifies  ‘robust’ Majorana zero modes. The color code is used to indicate the size of the topological gap, see
figure caption. In contrast, red and white shades mark regions where one of our criteria does not apply, the color encodes the size of the non-topological gap in this case. In the clean system, the red and green regions meet at the topological phase transition of the clean system (black dashed line). An exception is visible in the upper right corner of the plots in Fig.~\ref{fig:GapMapClean}, where topological Majorana edge modes exist but their localization length becomes so large that criterion (iii) is no longer passed. This region becomes smaller with increasing radius, since the superconducting pairing potential in units of the surface gap increases with increasing radius, $\Delta_\text{SC}/\Delta_\text{W}\approx0.06\frac{R_\text{W}}{10\,\text{nm}}$, and thus the localization length decreases in the clean system.

In Fig. \ref{fig:GapMap} we show how the result changes in the presence of disorder. We consider
 wires of radius 10, 15 and 20\,nm for three fixed disorder potentials with the impurity density of $10^{19}\,\text{cm}^{-3}$. For the thinnest wire, $R=10$\,nm, robust Majorana modes can be found in a large parameter regime. Also the topological gap is approximately as large as in the clean system, indicating that in-gap states do not play a major role.  Only close to the phase transition of the clean system disorder becomes important. 
 
At $R=15$\,nm we see more significant signatures of disorder. A substantial fraction of the phase diagram is now white, indicating the absence of robust Majorana modes. But by varying gate voltage and flux, one can easily find regions with robust Majorana zero modes according to our definition.
At $R=20\,$nm, regions with robust Majorana modes shrink further and become more scattered. Nevertheless, one can almost always find parameters where robust Majorana modes exist, as we will analyze in more detail below. 
\begin{figure}[t]
\centering
\includegraphics[width=8cm]{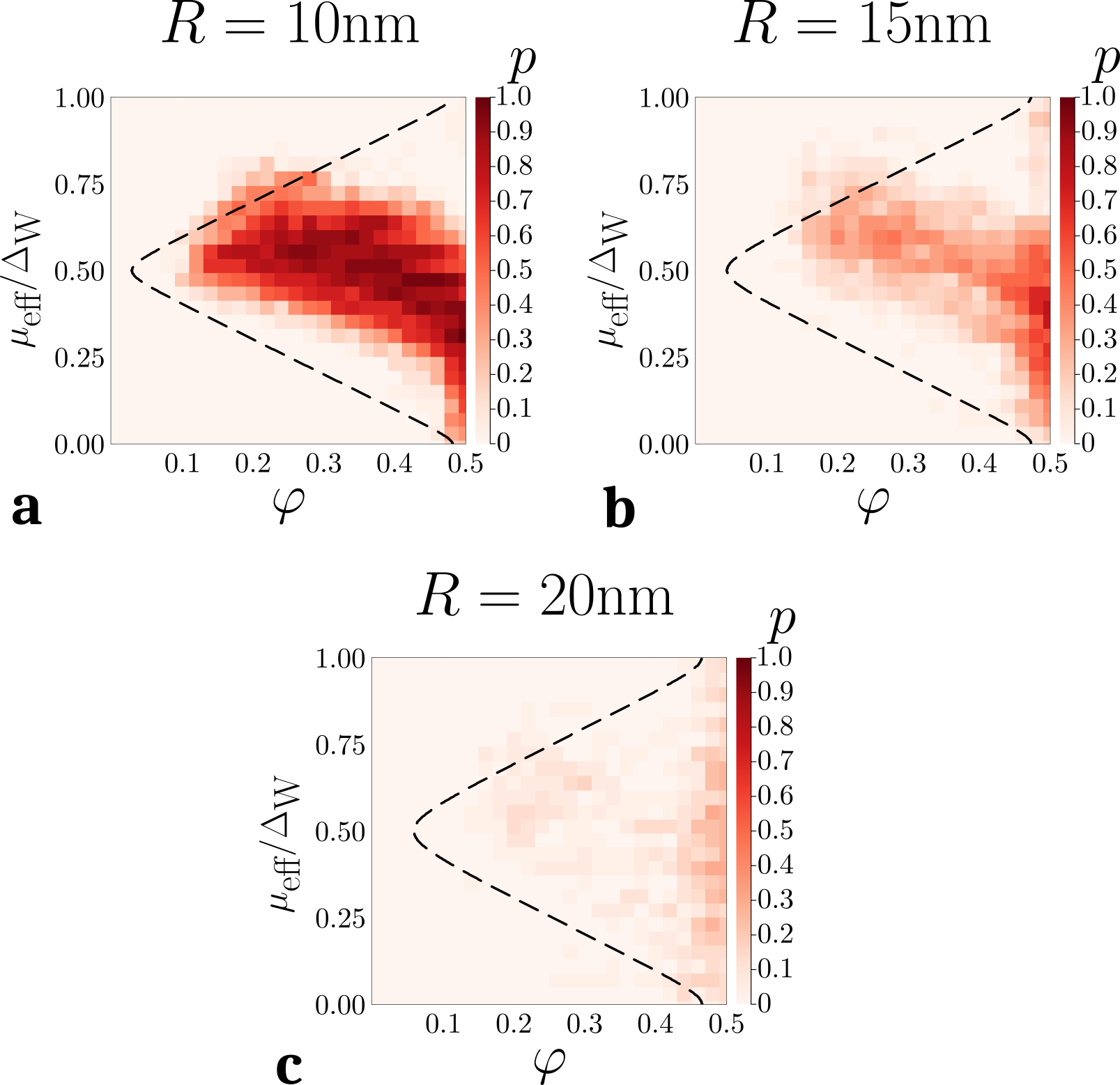}
\caption{\label{fig:topfracmaps} Fraction $p$ of samples, which show a robust Majorana zero mode for a given value of flux $\varphi$ and $\mu_\text{eff}$, i.e., without the fine-tuning or parameters (average over 30 samples, one such sample is shown in Fig.~\ref{fig:GapMap}).}
\end{figure}

To obtain a statistically relevant results, we have repeated the analysis sketched above for 30 disorder configurations for each of the three wire radii. Fig.~\ref{fig:topfracmaps} shows what fraction of the samples show a robust Majorana mode for a given parameter. This plot shows more clearly, that at $R=20$\,nm the disorder effects are substantial and without any fine-tuning the probability to find a robust Majorana is reduced.

\begin{figure}[t]\label{fig:gapvsdensity}
\centering
\includegraphics[width=8.7cm]{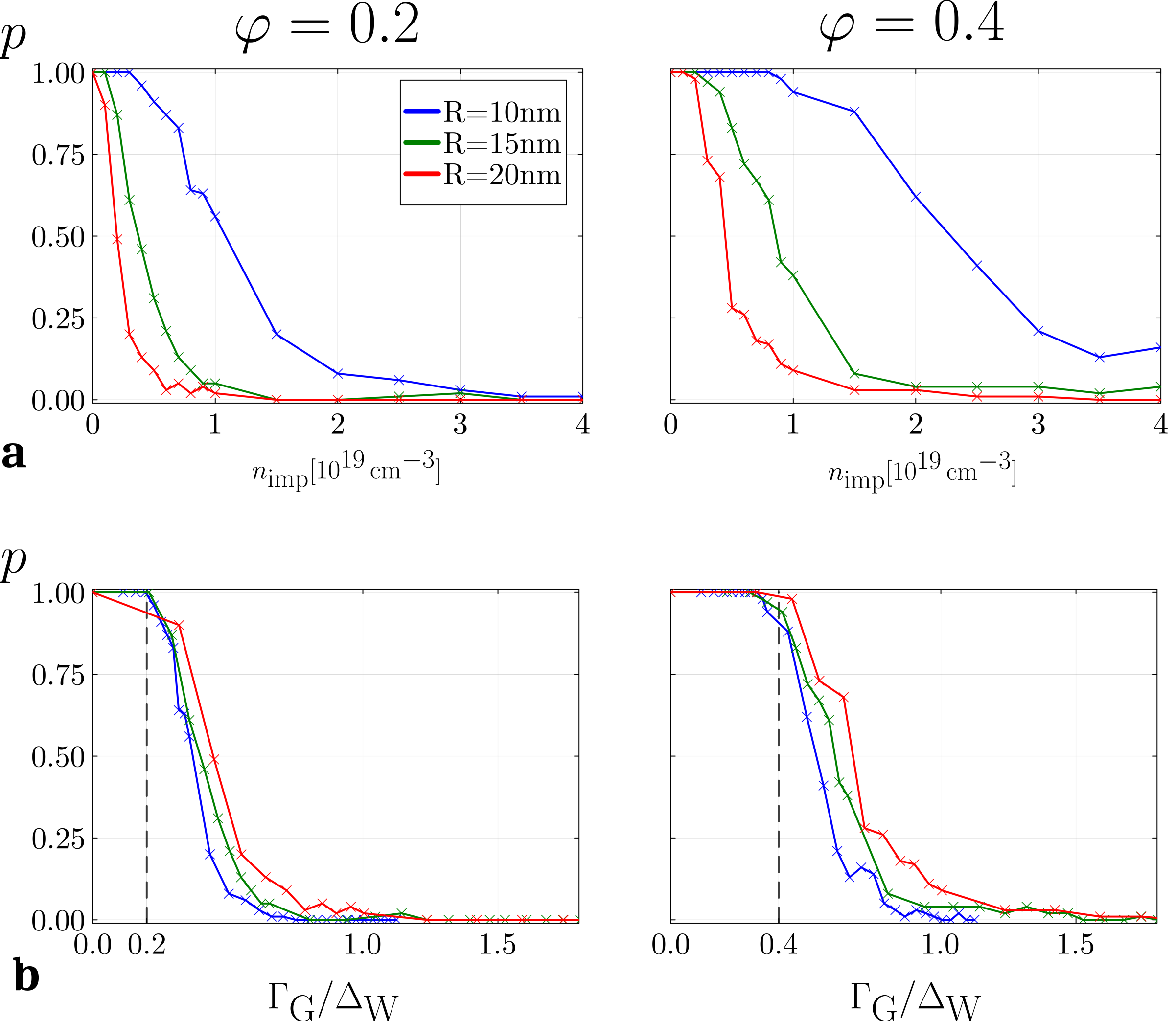}
\caption{\label{fig:p_robust_disorder} Influence of impurity density. Panel {\bf a}: Fraction $p$ of samples with a  robust Majorana zero mode  as a function of the impurity density for a fixed value of the gate voltage, and a fixed value of the flux $\varphi$. Note that much larger values of $p$ are obtained when flux and gate voltages are fine-tuned, see Fig.~\ref{fig:topfracsum}. Panel {\bf b}: The same data is shown as a function the disorder amplitude on the gate side of the device, see Fig.~\ref{fig:V_vs_n}. The plot suggests that defects located on the gate side of the device are most important, see Fig.~\ref{fig:sampleMajorana} for an example.
Parameters: $U_\text{G}\approx-1.15,\,-0.77,\,-0.55\,$V (such that $\mu_\text{eff}\approx \Delta_\text{W}/2$ in all cases) for $R=10,\,15,\,20,\,$nm }
\end{figure}

The density of charged impurities, $n_\text{imp}$,  is in general not known. As discussed above, $n_\text{imp}\approx 10^{19}\,\text{cm}^{-3}$, has been obtained from fits to experimental systems but one can expect that substantially cleaner -- or dirtier -- bulk-insulating samples can be obtained by varying details in the sample fabrication. Therefore, we show in Fig.~\ref{fig:p_robust_disorder}{\bf a} the fraction of robust Majorana modes as function of $n_\text{imp}$ for two values of the flux, $\varphi=0.2$ and $\varphi=0.4$, and for a gate voltage tuned to the optimal value $\mu_\text{eff}=\Delta_\text{W}/2$.
The larger the wire radius $R$, the cleaner the sample has to be to obtain robust Majorana states. Here, a central question is which parameter is responsible for the destruction of robust Majorana states. An increase of $R$ has two main consequences: (i) the characteristic size of the band gap of the TI-nanowire mini-bands, $\Delta_\text{W}=\hbar v_\text{F}/R$, is reduced. This affects  also the width of the topological phase as function of $\mu_\text{eff}$, see Fig.~\ref{fig:GapMapClean}.
(ii) The amplitude of the disorder potential on the gate side, $\Gamma_\text{G}$, increases, because the distance to the superconductor, responsible for screening, gets larger, see Eq.~\eqref{eq:Gamma_gate} and Fig.~\ref{fig:V_vs_n}{\bf a}.

In Fig.~\ref{fig:p_robust_disorder}{\bf b}, we plot the fraction of robust Majorana modes as function of $\Gamma_\text{G}/\Delta_\text{W}$. We obtain an approximate scaling collapse of the data for the three radii. For $\varphi=0.4$ the crossover occurs at $\Gamma_\text{G}/\Delta_\text{W}\approx 0.8$, while for $\varphi=0.2$, the crossover is located at $\Gamma_\text{G}/\Delta_\text{W}\approx 0.4$.

This has a simple qualitative explanation: for $\varphi= 0.4$ the width of the topological phase of the clean system as function of $\mu_\text{eff}$ is approximately twice as large compared to $\varphi=0.2$, see Fig.~\ref{fig:GapMapClean}. Therefore, we can also expect that it is more robust to random potential fluctuations.

For sufficiently small disorder, $\Gamma_\text{G}/\Delta_\text{W}\lesssim 0.4$ for $\varphi=0.4$ and $\Gamma_\text{G}/\Delta_\text{W}\lesssim 0.2$ for $\varphi=0.2$, we can neglect disorder effects completely.
Interpolating this result, $\Gamma_\text{G}/\Delta_\text{W}\lesssim \varphi$ (in the approximately linear region $2\Delta_\text{SC}/\Delta_\text{W}\le\varphi\le 0.45$) we obtain in 
combination with Eq.~\eqref{eq:Gamma_gate} the condition for negligible disorder effects
\begin{align}
    n_\text{imp}\lesssim 8\varphi^2\cdot  10^{19}\,\text{cm}^{-3} \left(\frac{10\,\text{nm}}{ R}\right)^3.\label{eq:critical_nimp}
\end{align}

Note that the prefactor is not universal but depends on the details of the model including the electromagnetic environment. 
We expect a weak $L$ dependence of the prefactor proportional to $\ln L/R$ due to the following argument.
A long wire probes roughly $L/R$ independent disorder configurations and thus the probability $p_r$ for a rare disorder configuration producing a low-energy Andreev bound state in the bulk should be $p_r \lesssim R/L$. As rare disorder potentials with an effective amplitude $V$ have an exponentially small probability, $p_r \sim e^{-(V/\Gamma_G)^2}$, the critical impurity concentration, $\propto \Gamma_G^2$, depends linearly on $\ln(L/R)$. A similar argument has previously been given by Huang and Shklovskii in  Ref.~\cite{Huang2021}. The same logarithmic correction, also affects the scaling collapse in Fig.~\ref{fig:p_robust_disorder}{\bf b} where wires with smaller radius (and thus more independent impurity configurations) are slightly more affected by the disorder. Our range of $R$ is, however, much too small to fit a $(\ln(L/R))^{1/2}$ dependence to the data.

\begin{figure}[t]
\centering
\includegraphics[width=8.50cm]{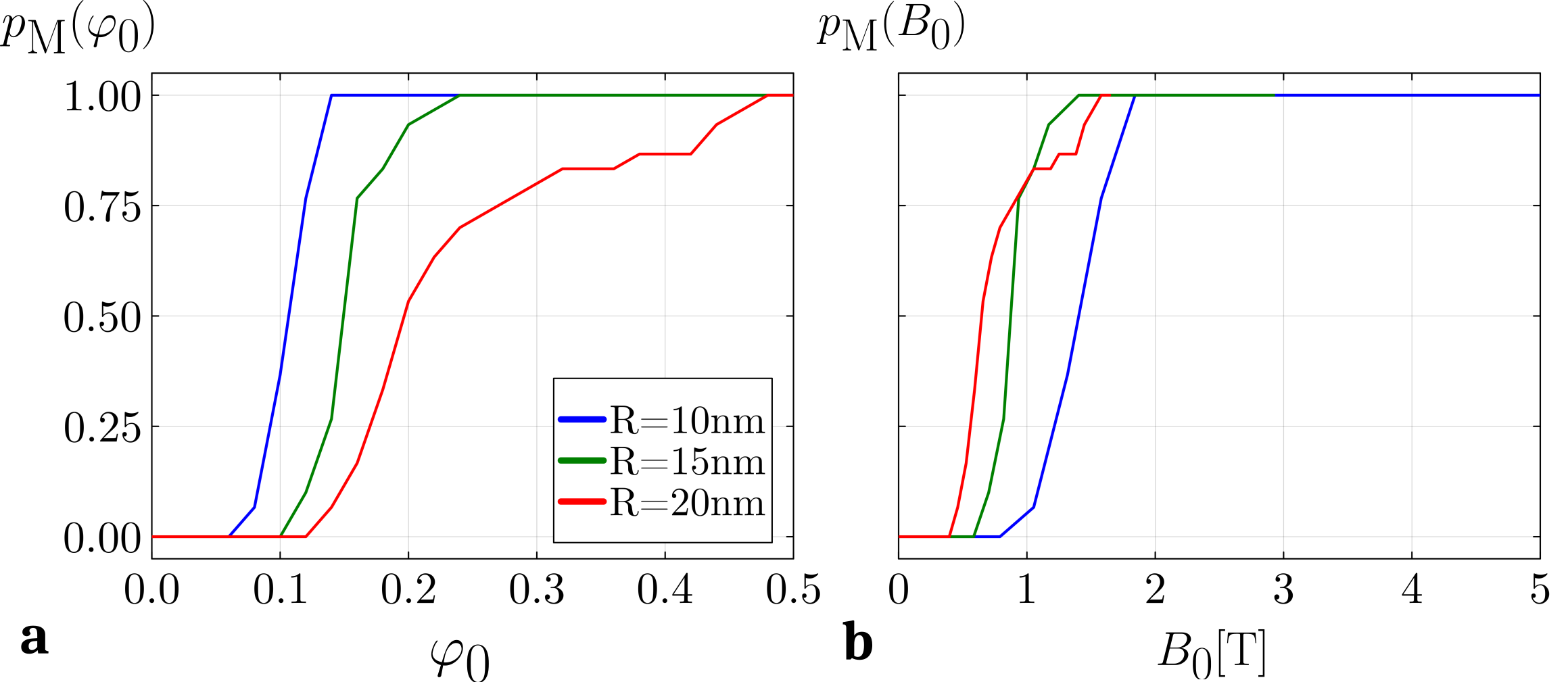}
\caption{Fraction of samples where robust Majorana modes can be found (by fine-tuning gate voltage and flux) for magnetic fluxes (panel {\bf a}) and fields (panel {\bf b}) in the range $\varphi<\varphi_0$ and $B<B_0$, respectively ($n_{\text{imp}}=10^{19}\,\text{cm}^{-3}$). Note that we only consider $\varphi_0\le 1/2$, as larger fields lead are detrimental for the formation of Majorana states, see Fig.~\ref{fig:phaseDiagramClean}. Thus the 
red and green line in panel {\bf b} terminate at $B_0\approx2.9\,$T and $B_0\approx 1.6\,$T for $R=15\,$nm and $R=20\,$nm, respectively. 
Thinner wires require larger magnetic fields to obtain robust Majorana modes.\label{fig:topfracsum}}
\end{figure}

The analysis given above does, however, paint a  too pessimistic picture.
Above, we were estimating under which conditions disorder become so small, that for one {\em fixed} value of gate voltage and field one obtains a robust Majorana state. For an experiment, one can (and should) ask instead the following question: can one find in a given sample robust Majorana modes by tuning the gate voltage and the magnetic field? 
We therefore define the following quantity: $p_\text{M}(\varphi_0)$ is the fraction of samples where one can find a robust Majorana for a flux $\varphi<\varphi_0$ by tuning both gate voltage and the external magnetic field. In Fig. \ref{fig:topfracsum}{\bf a}, we show $p_\text{M}(\varphi_0)$ for three wire radii. For $R=10$\,nm the function shows a very steep rise and jumps to $1$ for $\varphi_0\approx 0.14$. For larger radii, when disorder becomes more important, the curves rise more smoothly. For example, at $R=20$\,nm and for $\varphi_0=0.25$, in 73\% of the samples a robust Majorana mode can be found according to our definition of robustness. If one does not allow for fine-tuning of gate voltages and flux, one obtains for the same parameters only in roughly 10\% of the samples  a robust Majorana mode, see Fig.~\ref{fig:topfracmaps} and Fig.~\ref{fig:p_robust_disorder}. Consequently, this also means that Eq.~\eqref{eq:critical_nimp} provides a too strict limit for sample quality. Even with a  factor 3 or 4 times larger disorder density one can still find robust Majorana modes in almost all samples.

It is instructive, to plot the  data of  Fig.~\ref{fig:topfracsum}{\bf a} in a different way, not as a function of flux but as a function of magnetic field $B_0=\varphi_0/(\pi R^2)$, see Fig.~\ref{fig:topfracsum}{\bf b}. For the wire with a larger radius a substantially smaller magnetic fields are necessary to induce a sizable flux and thus robust Majorana states. The magnetic fields required to induce robust Majorana modes actually \textit{decreases} with increasing radius. As too large magnetic fields may suppress superconductivity directly in the parent superconductor --an effect not taken account  in our calculation-- it might be experimentally favorable to use wires with a larger radius, despite the fact that disorder effects are substantially bigger in this case  (assuming the same density of impurities).

\subsection{Discussion of neglected effects}\label{sec:unaccounted}

In our model, we have provided a semi-quantitative model of TI nanowires proximitized by a superconductors. Importantly, this included a model of the electrostatic environment and of the effects of charged impurities. When possible, we used ab-initio parameters to estimate, e.g., the position of the surface states and the size of proximity gaps. At the same time, our model is necessarily idealized and excludes a number of effects which we briefly discuss in the following.

First, our model focuses on surface states of a TI only, assuming that the bulk is gapped. Gated TI nanowires with bulk gaps have been reported by the Ando group in Ref.~\cite{Muenning2021}, where clear signatures of quantized TI surface states have been observed. In macroscopic samples of bulk TIs, charged impurities also induce charged regions in the bulk \cite{Skinner2012,Knispel2017,Thomas2017}, called bulk puddles. The effect of bulk puddles is, however, strongly suppressed in small nanowires where screening by surface states and the superconductor ensures that potential fluctuations remain small. In our simulations, the potential fluctuation are an order of magnitude smaller than the bulk gap of typical topological insulators ($\sim 300$\,meV). Therefore, it is justified to neglect the effect of bulk puddles.

Another major experimental question concerns the quality of the interface between TI and superconductor. Our modeling assumes perfect interfaces but defects (charged or not charged) at the interface may substantially weaken the proximity effect and provide another source of disorder, not accounted for in our model. Here, it would be useful to obtain further experimental input.

Another important effect not taken into account in our simulation are extra contact potential at the interface of the superconductor and the topological insulator. They arise whenever there is a mismatch of the work function (i.e., the energy needed to remove an electron) of the two materials. A large mismatch of work functions may, e.g., lead to band bending, a strong doping of the TI surface states and/or  the formation of extra non-topological surface bands, see, e.g., Refs.~\cite{Ruessmann2022,Chen2024}.

Ref.~\cite{Chen2024} studies this effect for a rotationally symmetric TI nanowires covered on all sides by a superconductor (full shell). Remarkably, the authors obtain large topological gaps even in regimes where the surface is strongly doped provided that the wires are sufficiently thick ($R\gtrsim 60$\,nm). Full-shell nanowires have the disadvantage that they cannot be controlled by external gates.
The physics of partially covered TI nanowires is expected to be qualitatively different as the contact potential will induce inhomogeneous potentials and Rashba coupling. This is an important topic for future studies.

Effects of contact potentials can be tuned by using  different superconducting materials and their alloys, and it would be interesting to explore this question in future ab-initio studies similar to Ref.~\cite{Ruessmann2022}. 

Finally,  we are not taking into account Zeemann effects from the magnetic field. In our model a term $g \mu_B B \sigma_x/2$ can fully be absorbed in a (tiny) shift of $\varphi$. As observed and discussed in Ref.~\cite{Taskin2017}, this neglects, e.g., possible disorder-induced inhomogeneities in $g$ factors.
Furthermore, the magnetic field can also suppress superconductivity directly but we argued above that this effect becomes smaller for wires with larger radius.

Furthermore, we would like to emphasize  that our study does not address the important questions how Majorana states can be experimentally detected and manipulated.

\section{Conclusions}

In this paper, we have focused on a single question: how does the presence of realistic levels of charged impurities affect the possibility to realized robust Majorana zero modes in half-shell TI nanowires. The result is promising: even for wires with a relatively large radius of 20\,nm and for a substantial disorder level of $10^{19}\,\text{cm}^{-3}$, one can find robust Majorana zero modes protected by a substantial topological gap, see Fig.~\ref{fig:GapMap}. Furthermore, if one is able to reduce the disorder level by an order of magnitude down to $10^{18}\,\text{cm}^{-3}$ (corresponding to about $10^3$ impurities per $\mu$m for a $20\,$nm-nanowire) then disorder effects due to charged impurities become completely irrelevant.

The main reason for this remarkable robustness to disorder is that the superconductor is very efficient in screening the charged impurities directly below the superconductor. By using a gate voltage with the right sign, one can ensure that the Majorana modes are pushed closer to the superconductor where the the proximity induced gap is largest and disorder levels are smallest.

Our study focuses on Coulomb disorder and does not take into account a number of effects arising from the detailed physical and chemical properties of the TI-superconductor interface. Extra defects at the interface, band bending effects due to work function mismatch, or details of the chemistry at the end of the wire can also strongly affect Majorana devices.

In our study, we have focused on TI nanowires. It is instructive to compare them to the most widely investigated alternative platform for Majorana qubits, semiconductor nanowires as investigated by Microsoft Quantum for which also a substantial amount of numerical modeling has been performed \cite{Woods2021,Microsoft2023}, including the effects of charged impurities. 
While not all details of the simulations on semiconductor nanowires have been published \cite{Kamien2023}, it is instructive to roughly compare the of the two systems.
The main advantage of the TI nanowires is that spin-orbit coupling is much larger in this case \cite{FuKane2008,Cook2011,Legg2021}. The typical order-of-magnitude of the Rashba coupling in Indium-based nanowires is of the order of $1\,\text{meV}$, to be compared to the corresponding scale in our system ($0.4 \,\Delta_\text{W}$ in Fig.~\ref{fig:phaseDiagramClean}), which is roughly an order of magnitude larger even for nanowires with a radius of $20\,$nm. While semiconductor nanowires rely on the Zeemann effect (enhanced by orbital effects) to open magnetic gaps, the TI nanowires use instead the Aharonov-Bohm effect arising from the magnetic flux. For a 20\,nm TI-nanowire, for example, about $1.5$\,T is sufficient to inject half a flux quantum, the optimum for realizing Majorana modes in terms of robustness against disorder. In Refs. \cite{Microsoft2023,Microsoft2025} topological gaps substantially below  100\,$\mu$eV  are reported both from model calculations for semiconductor nanowires \cite{Microsoft2023} and from experiments \cite{Microsoft2023,Microsoft2025}, about an order of magnitude smaller than in our calculations
(see Fig.~\ref{fig:GapMap} with $\Delta_\text{SC}\approx 1.6$\,meV taken from Ref.~\cite{Floetotto2018}). 

The comparison shows that TI nanowires appear to have an intrinsic advantage compared to the semiconductor nanowire as originally suggested by Cook and Franz \cite{Cook2011}. Our study shows, that this conclusion survives in the presence of realistic levels of charged impurities.

For future experimental and theoretical studies, a central question will be to explore further the role of the interface of superconductors and TI nanowires.

\begin{acknowledgments}
 We acknowledge useful discussions with Thomas B\"omerich, Jens Brede and, especially, with Henry Legg, Ella Nikodem, and Yoichi Ando. This work is funded by the
Deutsche Forschungsgemeinschaft (DFG, German Research
Foundation) under Germany’s Excellence Strategy-Cluster
of Excellence {\em Matter and Light for Quantum Computing}
(ML4Q) EXC 2004/1 - 390534769.
\end{acknowledgments}

\section*{Data availability}
All data and codes underlying this study are openly available via a Zenodo archive \cite{dataAvailabiliity}.

\bibliography{biblio.bib}
\clearpage

\appendix

\section{Numerical implementation}\label{AppendixA}
\subsection{Electromagnetic environment and the computation of Coulomb potentials}\label{AppendixA1}

\begin{figure}[t]
\centering
\includegraphics[width=5cm]{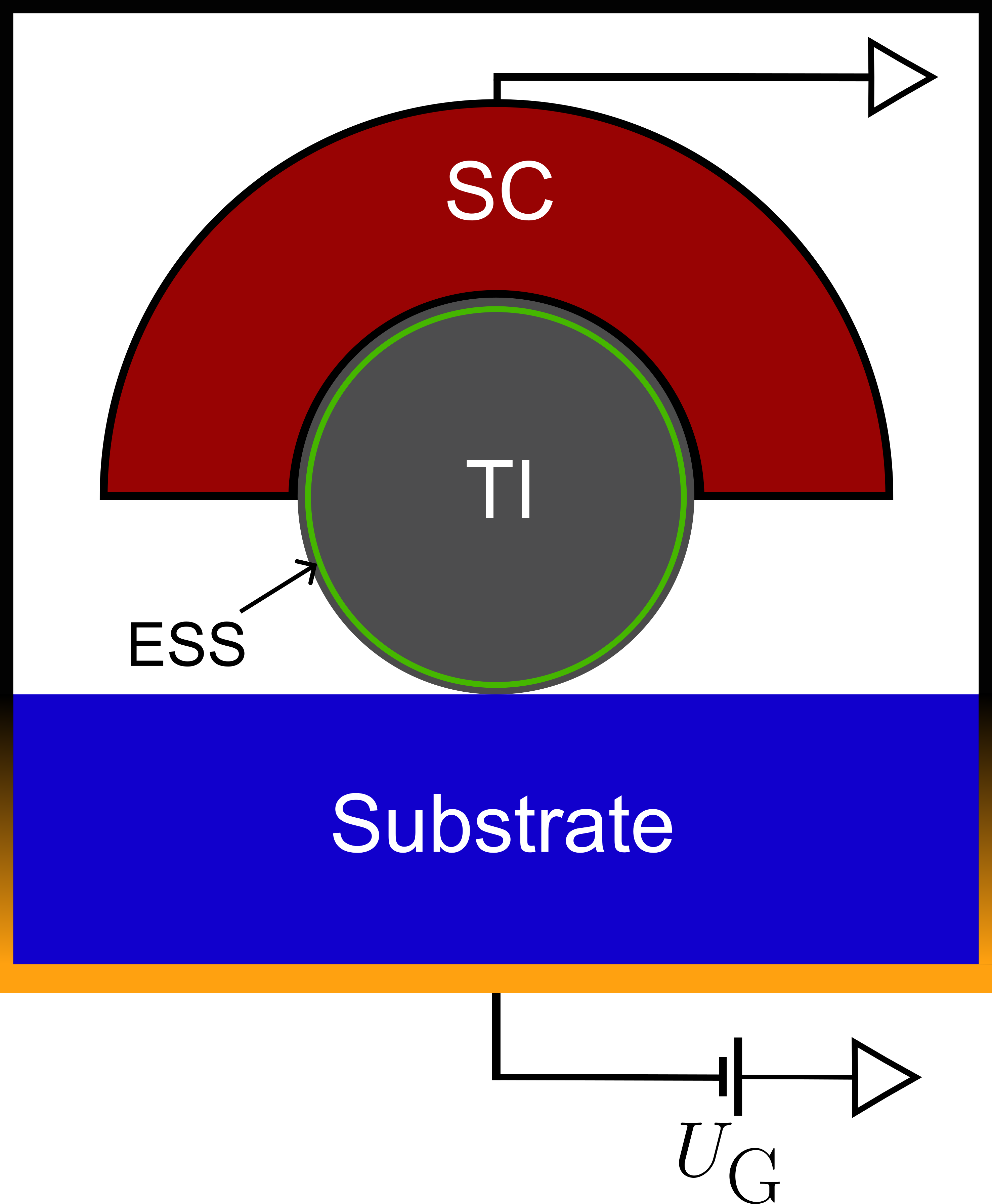}
\caption{\label{fig:NWNumerics}z-cut of the simulated box of size $5R_\text{W}\times5R_\text{W}\times L_\text{W}$, where $R_\text{W}$ is the wire radius and $L_\text{W}=1\,\mu$m is the length of the box. In $z$ direction, we use periodic boundary conditions. Black lines denote the boundary conditions $V=0$ (surface of the superconductor and upper part of the box). At the bottom of the box (orange), the gate voltage $U_\text{G}$ is applied and thus, we set $V=U_\text{G}$ as the boundary condition. In the shaded part of the outer box we linearly interpolate between $V=U_\text{G}$ and $V=0$.
Further parameters: substrate with width $1.5 R_\text{W}$, $\epsilon_\text{S}=4$, TI nanowire: radius $R_\text{W}$, $\epsilon_\text{W}=200$, superconductor width $R_\text{W}$, effective position of the electronic surface state (ESS): $5\,$\AA\ below the surface of the TI.}
\end{figure}

In this Appendix we describe details of the electrodynamic environment used in our calculations. Furthermore, we discuss how we solve the Poisson equation numerically to determine the relevant Coulomb potentials.
A sketch of our setup in shown in Fig.~\ref{fig:NWNumerics}, which also describes the parameters and boundary conditions used in our simulation. 
The potential of a single charge is modeled by the numerical solution of the Poisson equation for charges at random positions in the bulk. Most importantly, solving this problem provides a semi-realistic model for screening by the superconductor due to the image charge density induced there by the real charges inside the wire. While analytic solutions for slabs and cylinders of arbitrary dielectric constants are well established, the complex geometry of the nearby SC requires a numerical approach. Here, we employ the finite difference method with a few adaptions that enable us to simulate the 3 dimensional potential along the entire wire (and hence its surface). Since the system is, up to the impurity itself, assumed to be translationally invariant in z-direction, different momenta $k_z$ decouple and we can formulate the Poisson equation as 
\begin{align}\label{eq:Poisson}
   \left(k_z^2 \epsilon(x,y)- \nabla \epsilon(x,y)\nabla\right)\tilde{V}_{k_z}(x,y) = \frac{\delta(x-x_0)\delta(y-y_0)}{2\pi},
\end{align}
where $\nabla=\nabla_{x,y}$ is the two dimensional gradient and $\tilde{V}(x,y,k_z)$ the electrostatic potential Fourier-transformed along $z$-direction.

Effectively, this allows us to decompose the three dimensional Poisson problem into a set of two dimensional problems,  for which  the equation \begin{align}
    \mathbf{L}_{k_z}\Vec{\tilde{V}}_{k_z} = \Vec{Q}
    \end{align}
    has to be solved. We use a $N\times N$ square grid to discretize the 2d problem. We find that $N=500$ is sufficient for our purpose, which corresponds for $R=20$\,nm to a discretization of $d x=R/100=0.2\,$nm. 
     $\Vec{\Tilde{V}}$ and $\Vec{Q}$ are $N^2$-dimensional vectors encoding the two dimensional potential and charge, respectively. One may define a dielectric vector $\vec{\epsilon}$ encoding the dielectric environment in a similar manner. We choose a convention, where $\epsilon$ is defined on the plaquettes of our grid. As explained in the main text, we chose $\epsilon_\text{W}=200$ for the wire and $\epsilon_\text{S}=4$ for the substrate, typical values for Bi-based materials and $\text{SiO}_2$, respectively. The components of the linear Poisson operator on a grid $\mathbf{L}$ are most compactly written as
\begin{align*}
    \mathbf{L}_{i,i} &= \frac{\Vec{\epsilon}_i+\Vec{\epsilon}_{i-1}+\Vec{\epsilon}_{i-N}+\Vec{\epsilon}_{i-(N+1)}}{4}\cdot\left(k_z^2+\frac{4}{(d x)^2}\right) \\
    \mathbf{L}_{i,i+1} &=- \frac{\Vec{\epsilon}_i+\Vec{\epsilon}_{i-1}}{2 (dx)^2} \\
    \mathbf{L}_{i,i+N} &=- \frac{\Vec{\epsilon}_{i-N}+\Vec{\epsilon}_i}{2 (dx)^2}\\
    \mathbf{L}_{i,i-N} &=- \frac{\Vec{\epsilon}_{i-(N+1)}+\Vec{\epsilon}_{i-N}}{2 (dx)^2} \\
    \mathbf{L}_{i,i-1} &=- \frac{\Vec{\epsilon}_{i-1}+\Vec{\epsilon}_{i-(N+1)}}{2 (dx)^2}.
\end{align*}
Furthermore, we implement Dirichlet boundary conditions by fixing the potential $V_{i_b}=V_{i_b}^0$ for boundary sites with index $i_b$.
Technically, this is implemented by setting
 $\mathbf{L}_{i_b,j}=-\delta_{i_b,j}/(dx)^2$ and $Q_{i_b}=-V_{i_b}^0/(dx)^2$. Implementation of the superconductor as a metallic boundary condition and the gate at some finite potential therefore becomes straightforward. With this algorithm, we can  extract the Coulomb potential acting on the TI surface states. The finite penetration depth of the quantum states into the wire is taken into account by reading off the potential not on the surface of the wire but at a finite distance, which previous ab initio simulations have found to be $d\approx5$\,\AA \ \cite{Ruessmann2022}. 
 To obtain the potential from the charged impurities,
 we compute the Poisson equation for 1000 randomly selected two-dimensional coordinates $x_0,y_0$ inside the wire. We set $k_z=\frac{2 \pi}{L}n$ with $L=25\,R$ with $n\in \mathbb Z$, $|n|<400$. Thus, the maximal $k$ is given by $k_{\text{max}}=\frac{2 \pi}{R} 16 $, corresponding to a spatial resolution parallel to the wire of $dz=R/16=1.25$\,nm for $R=20\,$nm. The parameters have been optimized to minimize computational costs (and waiting time) while providing results with errors on the level of a few percent.

 For our results on potential fluctuations in a wire without a superconductor, we consider instead a wire in vacuum, i.e., in an environment with $\epsilon=1$ (without any gates and without the metallic box shown in Fig.~\ref{fig:NWNumerics}). In this limit, the Coulomb problem can be solved analytically in momentum space, which has been done in e.g. Refs~\cite{Teber2005,Cui2006}, and one may obtain the relevant Coulomb matrix elements from straightforward summations.

\subsection{Modeling of screening by surface states}\label{AppendixA2}

The algorithm described above captures the screening of the impurity potential by dielectric environment which includes the gates and the superconductor.

We have also investigated numerically the role of electron-electron interaction using a self-consistent Hartree approximation, described below. As discussed in the main text, we find that electron-electron interactions profoundly change screening in the {\em absence} of the superconductor. In the presence of the superconductor we find instead that the screening effects can be neglected. Therefore the algorithm described in this section is only used for Fig.~\ref{fig:samplePotential}{\bf b} and the computation of the renormalized gate, see the main text.

The self-consistent Hartree approximation, when evaluated in the presence of a given disorder potential, describes how potentials are modified due to electron-electron interactions due to screening. The screened total disorder potential is obtained from 
\begin{align}\label{screening eq}
    V_\text{tot}(\mathbf{r})=V_\text{ext}(\mathbf{r})+\int R d\theta' dz'\, V_{ee}(\theta,\theta',z-z')\rho_\text{ind}(\theta',z'),
\end{align}
where $V_{ee}(\theta,\theta',z-z')$ is the potential created by a single electron on the surface of the topological insulator with position $z'$ and azimuthal angle $\theta'$. Here $V_{ee}$ is computed from the algorithm described in Sec.~\ref{AppendixA1}.

 The change of the charge density due to the impurity potential, $\rho_\text{ind}(\theta,z)=\rho[H_V]-\rho[H_0]$, is obtained from the exact diagonalization of the  Hamiltonian 
\begin{align*}
    H_V =\sum_{k,k'}\sum_{l,l'} H_0(k,l)+V_\text{tot}(k-k',l-l').
\end{align*}
This allows to compute the charge density using
\begin{align}
    \rho(\theta,z)[V] = \sum_{E<\mu_0} |\psi_E(\theta,z)[V]|^2\label{eq:appHartreeChargeDensity}\\
    \rho_\text{ind}(\theta,z) = \rho(\theta,z)[V]-\rho(\theta,z)[V=0].\nonumber
\end{align}
In our experimental system $\mu_0$ is fixed by the contact with the superconductor.

While the equations described above are `only' based on self-consistent Hartree approximation, they fully capture the physics of screening on the level of the random-phase approximation (RPA), which is obtained by linearizing Eq.~\eqref{screening eq}. Importantly, it also includes non-linear screening effects which is essential due to the large size of the bare disorder potentials which can exceed the Fermi energy by an order of magnitude.
It also captures the physics of Friedel oscillations and the scattering from the corresponding Friedel potentials. Also the physics of Anderson localization is included as the single-electron problem is treated exactly. 
There are, however, also effects not covered by the approximation. Luttinger liquid theory predicts, for example, singular scattering from a single impurity. To estimate the importance of such effects, we use that screening is weakest on the gate side of the device, opposite to the superconductor and compute for two charges at $\theta=\theta'=0$ the dimensionless ratio \begin{align} 
\frac{V_{ee}(k=2\pi/R)}{\hbar v_\text{F}}\approx 0.007\ll 1,
\end{align} 
indicating that effective Luttinger liquid parameters will be very close to 1. The smallness of the number also explains why self-consistent screening effects from surface states can be neglected as discussed below.

To solve equation~\eqref{screening eq} self-consistently, we iterate the following algorithm until the desired accuracy is met:
\begin{enumerate}
    \item Calculate the n-th charge density $\rho^{(n)}(\theta,z)[V_\text{tot}]$ using Eq.~\eqref{eq:appHartreeChargeDensity}. This step starts at $\tilde{V}=V_\text{ext}$.
    \item Evaluate the right-hand side of Eq.~\eqref{screening eq} to obtain an intermediate $V_{\text{tot}}(\theta,z)$ 
    \item Calculate the next iterative potential as a linear combination of the new and old solution $V^{(n+1)}_\text{tot}=\kappa \tilde{V}+(1-\kappa)V^{(n)}_\text{tot}$ with a mixing constant $\kappa\leq 1$, which we have fixed to $\kappa=0.05$ to ensure convergence of the algorithm
\end{enumerate}

In the main text, see Fig.~\ref{fig:samplePotential}{\bf b}, we discuss the results of this algorithm for 
 a bare TI nanowire (i.e. without the superconductor/gate). In this case, electron-electron interactions are the main source of screening. The amplitude of the disorder potential is reduced by roughly one order of magnitude. For the investigated parameters the resulting disorder potential is of the order of the chemical potential (or of the band gaps).
 Furthermore, the correlation length along z-direction $\xi_z$ is strongly reduced. Using the fit $\langle V_\text{tot}(0)V_\text{tot}(z)\rangle\sim\text{exp}(-z/\xi)$, we obtain a correlation length of the order of the wavelength of electrons, $\xi\sim\lambda_F=2\pi/k_F (\mu_\text{eff})\approx\pi R$.

\begin{figure}[t]
\centering
\includegraphics[width=8cm]{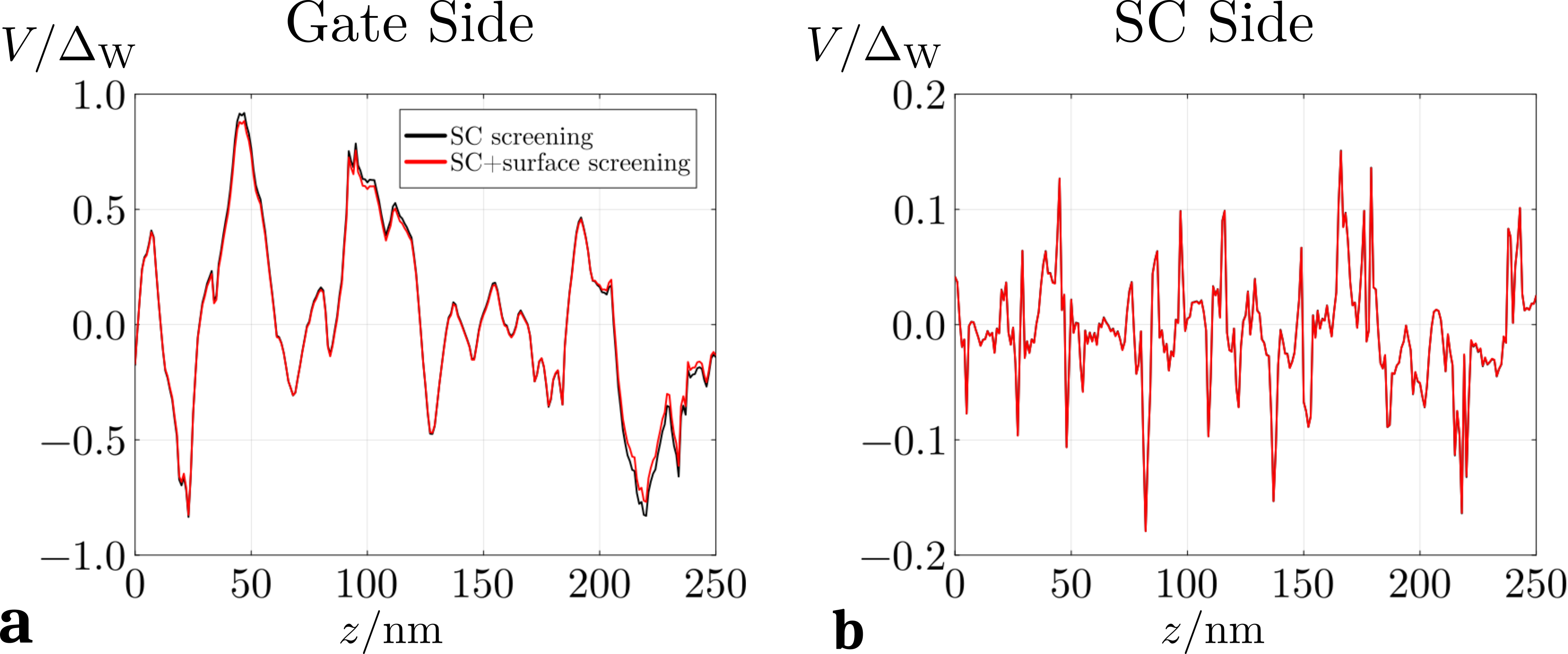}
\caption{\label{fig:AppendixScreeningComparison} Comparison between screening exclusively by the superconductor ($V_\text{ext}$, black curves) and superconductor+surface states ($V_\text{tot}$, red) for a wire with length $L=250\,$nm. Panel \textbf{a} shows the potential on the gate side, where the relative potential amplitude is reduced by roughly $5\,\%$ due to surface screening.
On the SC side, panel \textbf{b}, the effect is essentially invisible, on the level of $0.3\,\%$. Parameters: $R=10\,$nm, $\mu_0=0.5\,\Delta_\text{W}$, $U_\text{G}=0$.}
\end{figure}

 In the presence of the superconductor, however, the screening physics changes completely. Fig.~\ref{fig:AppendixScreeningComparison} compares a fully screened disorder potential $V_\text{tot}$ (red) with the potential $V_\text{ext}$, which is screened only by the superconductor in a wire with length $L=250\,$nm. The screening by surface electrons becomes negligible, reducing the potential amplitude by no more than 5\% on the gate side and less than 1\% on the SC side. 

 Furthermore, the correlation length of the disorder potential is reduced. Table~\ref{table:AppendixCorrelationLength} shows the correlation length along $z$ in units of the wire radius for wires of different radii.
 
 The superconductor affects the screening in two ways. First it reduces the initial potential $V_\text{ext}$ strongly, thus one enters the regime of linear screening. Second, it also reduces simultaneously the Coulomb potential of each electron, described by the term $V_{ee}$ in Eq.~\eqref{screening eq}.

\begin{table}[]
\begin{tabular}{|c|c|c|}
\hline
$\xi_z/R$ & Gate-Side & SC-Side \\ \hline
R=10nm    & 0.67      & 0.062   \\ \hline
R=15nm    & 0.62      & 0.053   \\ \hline
R=20nm    & 0.62      & 0.040   \\ \hline
\end{tabular}
\caption{\label{table:AppendixCorrelationLength}Correlation length $\xi_z$ along z-direction in units of the radius on  two opposite sides of the wire. On the gate side, $\xi_z/R$ is a constant of order $1$ consistent with the scaling arguments described in the main text of the paper. On the SC side the dipole approximation, Eq.~\eqref{eq:GammaSC} of the main text, predicts  $\xi_{z,\text{dip}}/R\approx0.075\frac{10\text{nm}}{R}$, which evaluates to $0.075, 0.05, 0.038$ in approximate agreement with the numerical results.}
\end{table}

\section{Surface Hamiltonian of a wire with variable radius}\label{Appendix:Curved}
Here, we provide a compact derivation of the Hamiltonian used to model the topological edge states. We loosely follow Ref.~\cite{Eguchi1980} to derive a Hamiltonian for a cylindrical surface with a $z$-dependent radius, $R(z)$, which is capable of accurately describing the ends of the wire (by sending $R(z)\rightarrow0$).\\
In flat 2+1 dimensional spacetime, the massless Dirac equation is $i\gamma^\mu\partial_\mu\,\psi=0$, where $\gamma^\mu$ are $2 \times 2$ Dirac gamma matrices along the cartesian $x^0,x^1,x^2$, with, e.g., $\gamma^0=i \mathbb{1}$, $\gamma^1=\sigma_x$, $\gamma^2=\sigma_y$. For our problem however, we have to generalize the Dirac equation to curved space, defined by coordinates $x^\mu$ and metric $g_{\mu\nu}$, while the time variable is not affected by the curvature. In the formulas below, we use Euclidean time, $x^0=it$, to be able to compare directly to Ref.~\cite{Zhang2010}, where the same convention is used and set $c=\hbar=1$, where $c$ is identified with the Fermi velocity of the surface state.
The generalized Dirac equation then reads
\begin{align}
    \gamma^\mu(\partial_\mu-\frac{i}{4}R_\mu^{ab}\sigma_{ab})\psi=0,
\end{align}
with $\gamma^\mu\gamma_\nu=\delta^\mu_{\nu}$. 
Here, $R_\mu^{ab}$ is the spin connection, which can be understood as the gauge field generated by local Lorentz transformations, where $\sigma_{ab}=-\sigma_{ba}=-\frac{i}{2}\left[\gamma_a,\gamma_b\right]$ are the corresponding generators. In the effectively two dimensional space we are working in, this is just $\sigma_{12}=\sigma_z$, the Pauli-z matrix. Obtaining the spin connection is more involved and the derivation for our explicit geometry is our first objective. In terms of the Christoffel symbols $\Gamma^\nu_{\ \,\sigma\mu}=\Gamma^\nu_{\ \,\mu\sigma}$ and the zweibein $e_\nu^{\ \,a}$ associated with $g_{\mu\nu}$, the spin connection is given by
\begin{align}
    R_\mu^{ab} &= e_\nu^{\ \,a}\Gamma^\nu_{\ \sigma\mu}e^{\sigma b}+e_\nu^{\ a}\partial_\mu e^{\nu b} \\
    \Gamma^\nu_{\ \sigma\mu}&=\frac{1}{2}g^{\nu\kappa}(\partial_\sigma g_{\mu\kappa}+\partial_\mu g_{\sigma\kappa}-\partial_\kappa g_{\sigma\mu})\\
    g_{\mu\nu} &= e_\mu^{\ a}e_\nu^{\ b}\eta_{ab}.
\end{align}
As usual, raising/lowering of greek and latin indices is done via multiplication with the metric $g$ and $\eta$, respectively. In our context, we can put the zweibein $e_\mu^{\ a}$ in perhaps more familiar terms as the set of orthonormal tangent vectors on the two dimensional surface (which consequently diagonalize the metric tensor). We can also restrict ourselves to working with a zweibein (as opposed to a dreibein in a general 2+1 dimensional spacetime) due to time being flat. This means that the Minkowski metric $\eta_{ab}$ is to be replaced by the Euclidean metric, $\eta=\mathbf 1$. For the same reason, all Christoffel symbols containing the time derivative, as well as the spin connection $R_0^{ab}$, are zero. With these definitions and simplifications, it is straightforward to find the spin connection for our geometry. In the following, we keep the z-dependence implicit and write $R(z)=R$ and the derivative $R'\coloneq \frac{dR}{dz}$.\\
We parametrize the surface by the coordinates $x^1=\theta$ and $x^2=z$, which relate to the Cartesian coordinates of the embedding three dimensional space as
\begin{align*}
x=R\,\text{cos}(\theta);\quad y=R\,\text{sin}(\theta);\quad z=z.
\end{align*}

The natural metric and zweibein of a cylindrical surface are well known, but obtain a correction to their second component due to the local curvature along z with space components
\begin{align*}
    g &= \begin{pmatrix}
R^2 & 0 \\
0 & 1+R'^2 
\end{pmatrix}\\
e^1=\begin{pmatrix}
R\\
0
\end{pmatrix}&\quad
e^2=\begin{pmatrix}
0\\
\sqrt{1+R'^2}
\end{pmatrix}.
\end{align*}
This gives the gamma matrices along the surface tangentials
\begin{align*}
    \gamma^\mu&=\gamma^ae^\mu_{\ a}\\
    \gamma^1=\frac{\sigma_x}{R},&\qquad\gamma^2=\frac{\sigma_y}{\sqrt{1+R'^2}},
\end{align*}
The non-zero Christoffel symbols are
\begin{align*}
    \Gamma^\theta_{\theta z}=\frac{R'}{R},\quad\Gamma^z_{\theta\theta}=-\frac{RR'}{1+R'^2},\quad\Gamma^z_{zz}=\frac{R'R''}{1+R'^2}.
\end{align*}
and the spin connections are given by
\begin{align*}
    R^{12}_\theta=\frac{R'}{\sqrt{1+R'^2}},\qquad R^{12}_z=0.
\end{align*}
Collecting all objects, we find the space component of the Dirac Hamiltonian to be
\begin{align}\label{eq:cylinder_Dirac}
    H_D = i\left(\sigma_x\frac{1}{R}\partial_\theta+\sigma_y\frac{1}{\sqrt{1+R'^2}}\left(\partial_z-\frac{iR'}{2R}\right)\right),
\end{align}\\
with boundary conditions $\Psi(\theta=2 \pi,z)=-\Psi(\theta=0,z)$ (arising from the $\pi$ rotation of the spinor when encircling the nanowire). The total length of the wire is $L+b$ with $b=0.25\,L$ with periodic boundary conditions in $z$ direction,
 $\Psi(\theta,-b/2)=\Psi(\theta,L+b/2)$.

Using this Hamiltonian, we model the `ends' of the wire by introducing a function $R(z)$, which interpolates from $R(z_\text{inside})=R_\text{W}$ to $R(z_\text{outside})=R_\text{min}\ll R_\text{W}$.
\begin{align}
    R(z)=\frac{R_\text{W}-R_\text{min}}{\exp((|z-L/2|-L/2)/R_\text{W})+1}+R_\text{min}.
\end{align} 
The chosen interpolating function (in analogy to the Fermi function) cuts off the radius exponentially on the scale of the radius. We use $R_\text{min}=10^{-2}R_\text{W}$. The region with very small $R$ obtains a huge band gap $\propto v/R$ which is larger than any other scale in our problem. Thus, the wire turns into a trivial band insulator in the region where $R(z)\ll R_\text{W}$ and is effectively cut off close to $z=0$ and $z=L$. At the same time, we are able to model the surface at the end of the wire, where the Majorana state is mainly located, see Fig. \ref{fig:sampleMajorana} of the main text.

For a magnetic field oriented parallel to the wire, the magnetic flux is given by $\pi R(z)^2 B$. We can model it by minimal coupling, i.e. by replacing \begin{align}
\partial_\theta \to \partial_\theta-\frac{B R(z)^2}{\Phi_0},
\end{align}
where $\Phi_0=\frac{2 \pi \hbar}{e}$ is the flux quantum. 

\section{Pfaffian invariant of the disordered topological insulator}\label{app:pfaffian}
One way to identify topological superconductors is the computation of a topological invariant based on the computation of Pfaffian \cite{Kitaev2001}. The computation of this invariant is, however, numerically expensive and therefore we have only used it for a simplified model with a shorter model length with constant radius, see below. 

For the computation of the Pfaffian invariant, one compares the many-particle wave function with periodic and antiperiodic boundary conditions. In the case of a topological insulator, the boundary conditions leads to a coupling of the Majorana modes at the end of the wire, $i t_b \gamma_L \gamma_R$, where the sign of $t_b$ is opposite for periodic and antiperiodic boundary conditions. Thus, the boundary condition controls the parity of the state, which is measured via the sign of the Pfaffian. This condition will, however, only work if the tunneling via the boundary $t_b$ is larger than the tunneling through the bulk of the sample which can be a limiting factor when using the topological invariant. 

For this reason, we have to choose a relatively narrow trivial region separating the two wire ends. We use $b=0.05\,L$ with $L\approx900\,$nm, where the total length of the wire is $L+b$. 
In the trivial region we set all potentials and the flux to zero, $\mu_0,\,V_\text{Gate},\,\varphi,\Delta_\text{SC}=0$, while keeping $R=const.$ This realizes a trivial insulator with gap $\hbar v_F/R$.

Within this model, the splitting of Majorana modes is determined (by construction) by their tunneling rate through the trivial region. We obtain a typical splitting of the order of  $|\Delta_\text{M}|\approx10^{-2}\Delta_\text{SC}$ for the chosen parameters.

According to the Pfaffian invariant a state is a topological superconductor if the parity of the the wavefunctions changes when one switches from periodic to antiperiodic boundaries. Our goal is to compare this criterion to the one introduced in Sec.~\ref{sec:topoDis} of the main text. More precisely, due to the increased Majorana splitting discussed above, we have to relax condition (i). We now demand $|\Delta_\text{M}|\leq5\cdot10^{-2}\Delta_\text{SC}$ (instead of $|\Delta_\text{M}|\leq 4\cdot10^{-3}\Delta_\text{SC}$) for the identification of Majorana modes. Similarly, the localization criterion (iii) has to be relaxed to 0.2 from the 0.3 in Eq.~\ref{eq:MZMcriterion3} of the main text.

Below, we will show that even this relaxed condition is much more strict than the Pfaffian invariant in the identification of robust Majorana states. 
\begin{figure}[t]
\centering
\vspace*{2mm}\includegraphics[width=8cm]{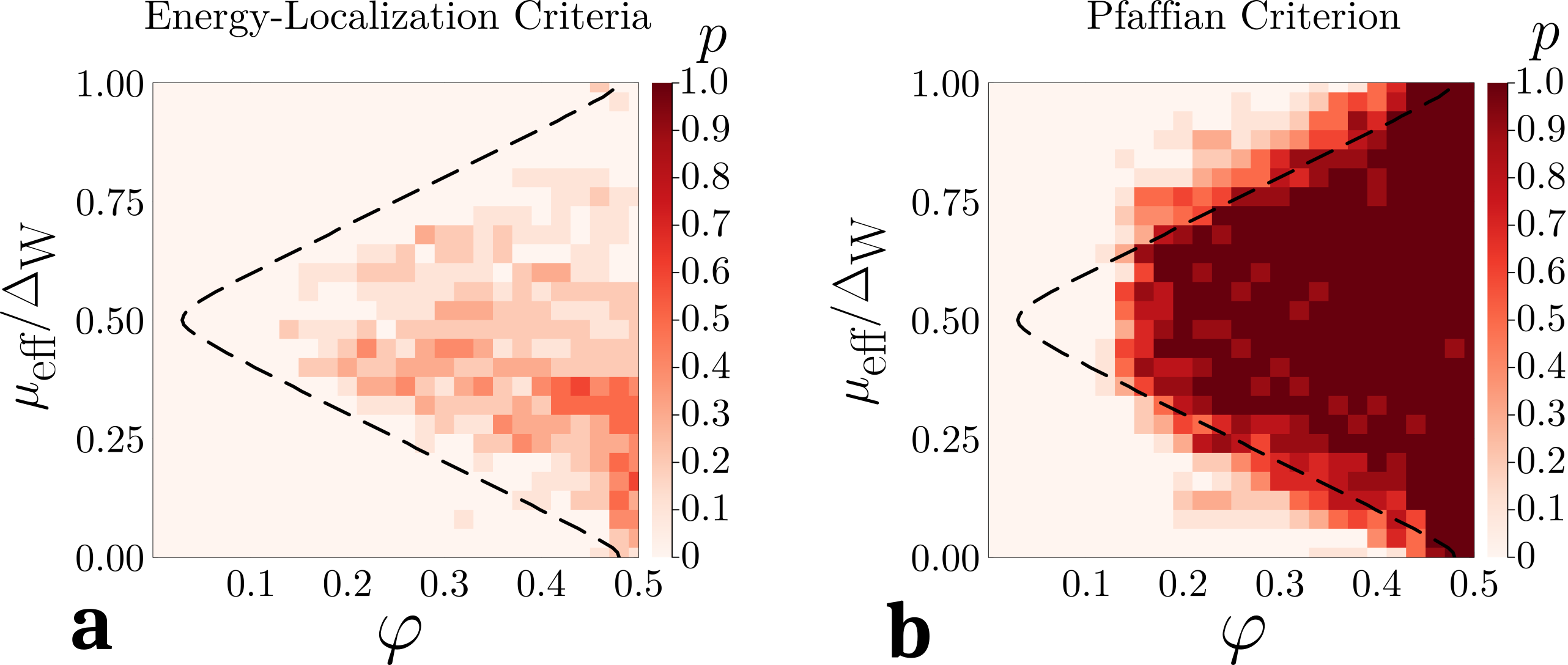}
\caption{\label{fig:AppendixE-L vs Paffian R=15nm} Panel \textbf{a}: Fraction of samples passing the three criteria of our algorithm. Panel \textbf{b}: Fraction of samples that are identified as topologically non-trivial by the Pfaffian invariant. It is apparent, that the invariant is a much more lenient criterion, only producing a negative in $0.1\%$ of the investigated samples which were identified as positive in our algorithm. Parameters used for this plot: Averaged data over 10 samples with $R=15\,$nm and an impurity density $n_\text{imp}=10^{19}\,\text{cm}^{-3}$.}
\end{figure}

In Fig.~\ref{fig:AppendixE-L vs Paffian R=15nm} we show the fraction $p$ of samples that show robust Majorana modes for the given parameters $\varphi$ and $\mu_\text{eff}$, analogous to Fig.~\ref{fig:topfracmaps} of the main text. In Panel \textbf{a}, we define the notion of 'robust' via our conditions on energy and localization. Panel \textbf{b} shows for the same samples the topology obtained by the Pfaffian invariant. It is apparent that the Pfaffian produces vastly more positives than our (relaxed) scheme. 

Only in  $0.1\%$ of the total investigated data points or $1.8\%$ of the identified robust Majorana modes (8 out of 449 positives), we obtain a situation where the relaxed variant of our scheme  identifies a `robust Majorana' which is not confirmed by the Pfaffian invariant. We believe that this -- very small -- number of false positives is reduced even further when one restores the original, much more strict condition on $\Delta_\text{M}$.

The analysis presented above depends substantially on the width $b$ of the trivial insulator. For small $b$ the Pfaffian conditions identifies more cases as topological than 
for large $b$. For our model and in comparison to the criteria used in our paper, we conclude that it does not offer useful extra information.

\end{document}